\documentclass[prd,aps,twocolumn,reprint,nofootinbib,preprintnumbers]{revtex4-2}

\usepackage{amsmath,amssymb}
\usepackage{color}
\usepackage{graphicx}
\usepackage{hyperref}
\hypersetup{colorlinks, citecolor=blue, linkcolor=black, urlcolor=blue}

\usepackage[american]{babel}
\usepackage{microtype}

\newcommand{\Br}{\mathrm{Br}}
\newcommand{\mub}{\mu\mathrm{b}}
\newcommand{\eV}{\text{e\kern-0.15ex V}}
\newcommand{\keV}{\mathrm{k}\eV}
\newcommand{\MeV}{\mathrm{M}\eV}
\newcommand{\GeV}{\mathrm{G}\eV}

\begin{document}

\title{Probing Invisible Vector Meson Decays with NA64 and LDMX}
\preprint{SLAC-PUB-17635}

\author{Philip Schuster}
\email{schuster@slac.stanford.edu}
\affiliation{SLAC National Accelerator Laboratory, 2575 Sand Hill Road, Menlo Park, CA 94025, USA}
\author{Natalia Toro}
\email{ntoro@slac.stanford.edu}
\affiliation{SLAC National Accelerator Laboratory, 2575 Sand Hill Road, Menlo Park, CA 94025, USA}
\author{Kevin Zhou}
\email{knzhou@stanford.edu}
\affiliation{SLAC National Accelerator Laboratory, 2575 Sand Hill Road, Menlo Park, CA 94025, USA}

\begin{abstract}
Electron beam fixed target experiments such as NA64 and LDMX use missing energy-momentum to detect the production of dark matter and other long-lived states. The most studied production mechanism is dark Bremsstrahlung through a vector mediator. In this work, we explore a complementary source of missing energy-momentum signals: Bremsstrahlung photons can convert to hard vector mesons in exclusive photoproduction processes, which then decay to dark matter or other invisible particles, such as neutrinos. We find that existing NA64 data can improve the leading constraints on invisible light vector meson decays, while a future run of LDMX could improve them by up to $5$ orders of magnitude. For the examples of a dark photon and a $U(1)_B$ gauge boson mediator, accounting for meson decays substantially enhances these experiments' sensitivity, especially to thermal relic dark matter of mass $m_\chi \gtrsim 0.1 \, \GeV$. 
\end{abstract}

\maketitle

\section{Introduction}

Light dark matter (DM) in the sub-GeV mass range has received a surge of interest over the last decade, triggered by its potential to explain several direct and indirect detection anomalies~\cite{Boehm:2003hm,Fayet:2004bw,Gunion:2005rw}, and more generally by its viability as a WIMP-like thermal relic in simple dark sector models. Such models generically predict meson decays with missing energy, and several works explored flavor violating decays of $B$ mesons, $D$ mesons, and kaons~\cite{Dreiner:2009er,Badin:2010uh,Kamenik:2011vy}, and invisible and radiative decays of light flavorless mesons~\cite{Fayet:2006sp,Dreiner:2009er} and heavy quarkonia~\cite{McElrath:2005bp,Fayet:2007ua,McKeen:2009rm,Yeghiyan:2009xc}. More recently, invisible and radiative decays of light flavorless mesons~\cite{Darme:2020ral,Bertuzzo:2020rzo} and heavy quarkonia~\cite{Fernandez:2014eja,Fernandez:2015klv,Hazard:2016fnc,Bertuzzo:2017lwt,Li:2021phq} into DM have been reconsidered from an effective field theory perspective. 

Motivated by these predictions and others, flavor factories have set limits on invisible meson decay, as we show in Table~\ref{tab:branching_rates}. Invisible decays of the light mesons $\omega$, $\phi$, $\eta$, and $\eta'$ have been searched for at BES~\cite{BESIII:2018bec,BESIII:2012nen}, and the leading constraints on the invisible decay of heavy quarkonia $J/\psi$ and $\Upsilon$ have been set at BaBar~\cite{BaBar:2013npw,BaBar:2009gco}. Recently, NA62~\cite{NA62:2020pwi} has set a stringent limit on invisible decays of $\pi^0$ mesons. 

\begin{table*}
\centering
\begin{tabular}{c|cccc}
 & $\Br(V,M \to \text{inv})$ & $\Br(V,M \to \nu \bar{\nu})$ & $\Br(V,M \to \gamma + X_{\text{inv}})$ & $\Br(V,M \to \gamma \nu \bar{\nu})$ \\ \hline
$\rho^0$ & -- & $2.4 \times 10^{-13}$~\cite{Gao:2018seg} & -- & unknown \\
$\omega$ & $< 7 \times 10^{-5}$~\cite{BESIII:2018bec} & $2.8 \times 10^{-13}$~\cite{Gao:2018seg} & -- & unknown \\
$\phi$ & $< 1.7 \times 10^{-4}$~\cite{BESIII:2018bec} & $1.7 \times 10^{-11}$~\cite{Gao:2018seg} & -- & unknown \\
$J/\psi (1S)$ & $< 7 \times 10^{-4}$~\cite{BaBar:2013npw} & $2.7 \times 10^{-8}$~\cite{Chang:1997tq} & $< 1.7 \times 10^{-6}$~\cite{BESIII:2020sdo} & $7 \times 10^{-11}$~\cite{Gao:2014yga} \\
$\Upsilon (1S)$ & $< 3 \times 10^{-4}$~\cite{BaBar:2009gco} & $1.0 \times 10^{-5}$~\cite{Chang:1997tq} & $< 4.5 \times 10^{-6}$~\cite{BaBar:2010eww} & $2.5 \times 10^{-9}$~\cite{Fernandez:2015klv} \\ \hline
$\pi^0$ & $< 4.4 \times 10^{-9}$~\cite{NA62:2020pwi} & see caption & $< 1.9 \times 10^{-7}$~\cite{NA62:2019meo} & $2 \times 10^{-18}$~\cite{Arnellos:1981bk} \\
$\eta$ & $< 1.0 \times 10^{-4}$~\cite{BESIII:2012nen} & see caption & $\lesssim 5 \times 10^{-4}$~\cite{CrystalBarrel:1994zpx} & $\sim 2 \times 10^{-15}$~\cite{Arnellos:1981bk} \\
$\eta'$ & $< 6 \times 10^{-4}$~\cite{BESIII:2012nen} & see caption & $\lesssim 2 \times 10^{-6}$~\cite{CrystalBarrel:1994zpx} & $\sim 2 \times 10^{-14}$~\cite{Arnellos:1981bk}
\end{tabular}
\caption{Summary table for invisible and radiative decays of flavorless vector mesons $V$ and pseudoscalar mesons $M$. Most experimental bounds are as in Ref.~\cite{ParticleDataGroup:2020ssz}, except for invisible $\pi^0$ decay and radiative $\eta$ and $\eta'$ decay. The experimental bounds on invisible decays tag decays of a heavier meson and search for missing mass corresponding to the given meson, while those for radiative decays search for missing mass from an invisibly decaying $X$. In the Standard Model, these processes occur through decays to neutrinos. Note that for the pseudoscalar mesons, decays to two neutrinos are proportional to $m_\nu^2$ because of helicity suppression. Thus, decays to four neutrinos may dominate, but they are also extremely rare~\cite{Gao:2018seg}, being suppressed by $(G_F m_M^2)^4$.}
\label{tab:branching_rates}
\end{table*}

In this paper, we describe a new method for detecting invisible meson decay. Existing searches tag the invisibly decaying meson by producing it through the decay of a heavier meson. By contrast, missing energy/momentum experiments such as NA64~\cite{Andreas:2013lya,Bernhard:2020vca} and LDMX~\cite{LDMX:2018cma,LDMX:2019gvz} are sensitive to {\it any} process in which a beam electron transfers most of its energy to invisible particles, leading to a missing energy signal with no accompanying penetrating particles. The exclusive production and invisible decay of an energetic meson contributes to this inclusive missing energy signal. As a result, such experiments may be used to simultaneously set limits on the invisible branching ratios of all kinematically accessible mesons, potentially strengthening existing constraints by orders of magnitude. These electron beam experiments are conventionally interpreted as probes of the DM coupling to electrons, but as we will see, their sensitivity to invisible meson decays also offers a powerful probe of the DM or light mediator couplings to quarks. 

We will consider dark sector models where the DM particle $\chi$ interacts with a vector mediator $A'$ which in turn interacts weakly with quarks, focusing on the well-motivated examples of a kinetically mixed dark photon~\cite{Okun:1982xi,Holdom:1985ag} and a $U(1)_B$ gauge boson~\cite{Carone:1994aa,Carone:1995pu,FileviezPerez:2010gw,Graesser:2011vj}. (For reviews of dark sectors and dark photons, see Refs.~\cite{Essig:2013lka,Alexander:2016aln,Fabbrichesi:2020wbt,Filippi:2020kii}.) Such models can be probed by the invisible decays of flavorless vector mesons. This signature is particularly promising because photons impinging on nuclei can efficiently convert into these mesons, through exclusive forward photoproduction reactions that transfer little energy to the recoiling nucleus or nucleon. Our work thus complements Ref.~\cite{Gninenko:2014sxa}, which focuses on invisible decays of light pseudoscalar mesons. 

\begin{figure}
\includegraphics[width=0.9\columnwidth]{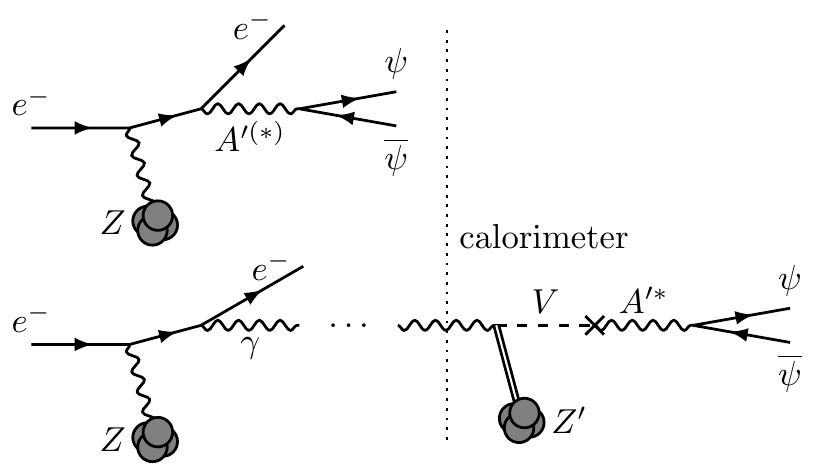}
\caption{Schematic depiction of the DM signal at LDMX from $A'$ Bremsstrahlung (top) and invisible vector meson decay (bottom). In the former, DM is produced through an on- or off-shell $A'$ in the target. In the latter, a hard photon is produced in the target, and converts to a vector meson $V$ in an exclusive photoproduction process in the calorimeter. The vector meson then decays invisibly to DM via mixing with the $A'$. 
}
\label{fig:meson_cartoon}
\end{figure}

A simple estimate demonstrates the potential of our approach. At NA64 or LDMX, the sequence of events that leads to a missing energy/momentum signal from invisible vector meson decay is shown in Fig.~\ref{fig:meson_cartoon}. The expected yield of the vector meson $V$ through exclusive photoproduction is $N_V = N_e f_{\text{brem}} p_V$, where $N_e$ is the number of electrons on target, $f_{\text{brem}}$ is the fraction that produce a hard Bremsstrahlung photon, and $p_V$ is the probability the photon undergoes an exclusive photoproduction process. 

\begin{figure}
\includegraphics[width=\columnwidth]{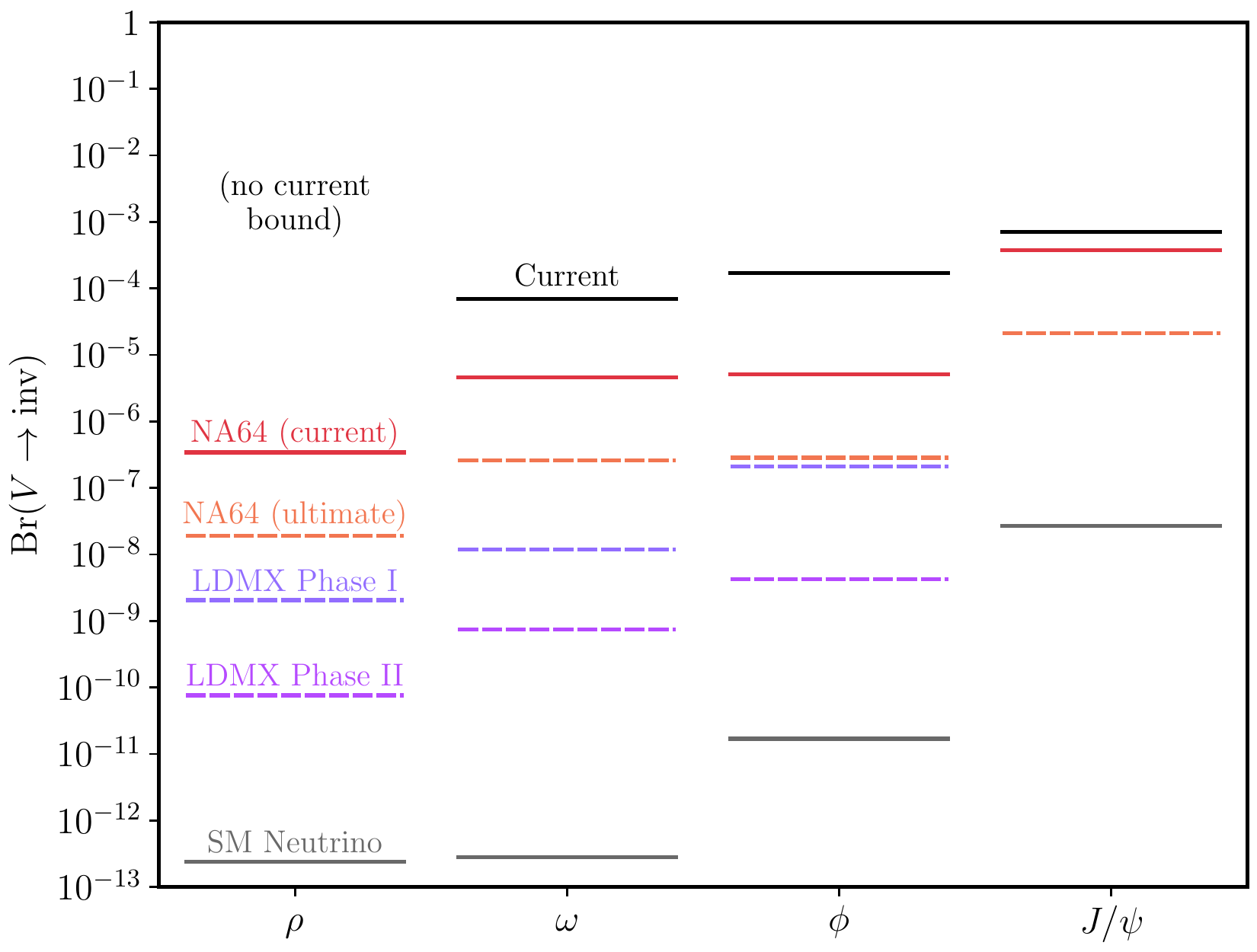}
\caption{Bounds on invisible meson decay, summarizing information from Tables~\ref{tab:branching_rates} and~\ref{tab:yields}. We show the best current bound, our projected 90\% C.L.~exclusions for four experimental benchmarks (assuming zero background events), and the invisible branching ratio within the SM due to decays to neutrinos.}
\label{fig:branching_summary}
\end{figure}

Most photons initiate an electromagnetic shower through a photon-conversion process, with cross section $\sigma_{\gamma N \to e^+ e^- N} \simeq 7 m_N / 9 X_0$, where $m_N$ is the mass of the nucleus and $X_0$ is the radiation length. The photoproduction cross section is $\sigma_{\gamma N \to V N} = f_{\text{nuc}}^V A \sigma_0^V$, where $\sigma_0^V$ is the cross section for exclusive photoproduction on a single nucleon, and $f_{\text{nuc}}^V$ is an order-one correction factor. Thus, 
\begin{equation} \label{eq:prob_estimate}
p_V \simeq \frac{9}{7} \frac{\sigma_0^V X_0 f_{\text{nuc}}^V}{m_p} = 10^{-5} \frac{X_0}{12.86 \ \text{g}/\text{cm}^2} \frac{\sigma_0^V}{1 \ \mub} \, \frac{f_{\text{nuc}}^V}{1.0}
\end{equation}
where we have normalized to the radiation length for copper. Typically $\sigma_0^V$ is on the order of $1 \ \mub$, so that given the LDMX Phase II design parameters $N_e = 10^{16}$ and $f_{\text{brem}} = 0.03$, we expect meson yields on the order of $10^9$ to $10^{10}$. This leads to the strong projected bounds on invisible vector meson decay shown in Fig.~\ref{fig:branching_summary}. As we will see, at high $m_{A'}$, the corresponding sensitivity to dark sector models exceeds that due to $A'$ Bremsstrahlung, largely because the latter is parametrically suppressed by $(m_e / m_{A'})^2$.

The rest of the paper is structured as follows. In section~\ref{sec:expt}, we describe in greater detail how invisible meson decay can give rise to missing energy/momentum signals at NA64 and LDMX. In section~\ref{sec:photoprod}, we estimate the exclusive photoproduction yields of the relevant vector mesons, reserving details for the appendix. We calculate the invisible branching ratios in the dark photon and $U(1)_B$ models in section~\ref{sec:ratios}, and show the resulting projected constraints in section~\ref{sec:results}. We conclude by discussing potential future directions, such as experimental studies and applications to neutrino physics, in section~\ref{sec:disc}.

\section{Missing Energy/Momentum Experiments}
\label{sec:expt}

Fixed target experiments have emerged as a powerful probe of light dark sectors~\cite{Bjorken:2009mm,Batell:2009di,Reece:2009un}. In this paper, we focus on the missing energy approach~\cite{Gninenko:2016kpg}, exemplified by NA64, and the missing momentum approach~\cite{Izaguirre:2014bca}, exemplified by the proposed LDMX experiment. In both cases, individual electrons from a low-intensity electron beam are tagged and directed at a target. Dark matter production through $A'$ Bremsstrahlung, shown at the top of Fig.~\ref{fig:meson_cartoon}, leads to an observed final state consisting solely of a much lower-energy (and transversely deflected) recoil electron, with the rest of the energy carried by the produced DM particles, which pass through the detector without interacting. These events are identifiable with order-one efficiency by measuring the electron's energy loss with downstream tracking and/or calorimeters, together with the absence of other detected particles that could have carried the energy away. Missing energy/momentum experiments must measure the detector response to one electron at a time, which limits their event yield. Nonetheless, they can match or exceed the sensitivity of much higher-luminosity beam dump experiments to weakly coupled light DM, as beam dumps are only sensitive to the small fraction of DM production events where the DM rescatters in a downstream detector. 

In the missing energy approach, the target is the front of the electromagnetic calorimeter itself. By contrast, in the missing momentum approach, DM production occurs in a thin target separated from the electromagnetic calorimeter, allowing the electron to be subsequently deflected and tracked downstream of the target. This enables a higher degree of background rejection, as well as in situ background measurements that would lend credence to any claimed discovery. 

In both experimental approaches, electrons often produce hard Bremsstrahlung photons in the target which carry away the majority of their energy, and such events must be rejected extremely reliably. A key potential source of background is the case where the hard photon initiates a hadronic shower through an exclusive photoproduction process, such as $\gamma p \to \pi^+ n$ or $\gamma N \to N K_S K_L$. Reliably rejecting these photonuclear reactions is an important design driver for the downstream electromagnetic and hadronic calorimeters, which has been studied in detail for LDMX in Refs.~\cite{LDMX:2018cma,LDMX:2019gvz}.

The main point of this paper is that the exclusive photoproduction of vector mesons, $\gamma N \to V N$, can also be an important source of \textit{real} missing energy/momentum signals. These vector mesons carry almost all of the original photon's energy, and decay well before directly interacting with any other nuclei in the calorimeters. If they decay to invisible final states such as DM, as shown at the bottom of Fig.~\ref{fig:meson_cartoon}, then the entire process leaves no trace of the original photon besides the recoil energy of the nucleus or nucleon. Therefore, it is crucial to understand the efficiency with which these recoils survive vetoes used by NA64 and LDMX to reject Standard Model backgrounds. While this is ultimately an experimental question, we will argue below that these survival probabilities should be of order one. 

As discussed in the following section, meson production proceeds through both coherent and incoherent processes. Coherent production involves scattering off an entire nucleus, and is peaked at very low momentum transfer $q \lesssim 1/r_{\text{nuc}}$. Even for a light target nucleus such as silicon, this corresponds to recoil kinetic energies below $100 \ \keV$. Such energy depositions are unobservably small, especially given the likelihood that the nucleus would never even reach active material of the calorimeter. Thus, coherent meson photoproduction produces an unambiguous missing energy signal. It accounts for most of the light meson yield at NA64, and about half of the meson yield at LDMX. However, it is form factor suppressed for heavy mesons, such as $\phi$ mesons at LDMX and $J/\psi$'s at NA64. 

In these cases, incoherent production dominates; it leads to recoils off individual nucleons, with characteristic momentum transfer $\sim 500 \ \MeV$. Therefore, the nucleons receive a typical kinetic energy $\sim 100\ \MeV$, and recoil at wide angles of $50^\circ$ to $70^\circ$ from the beamline. These energies are near the sensitivity limits of the detectors. 

For example, a proton with $50 \ \MeV$ kinetic energy at these angles would stop within one tungsten absorber layer of the LDMX ECal, and thus could be completely undetectable. By contrast, a $200 \ \MeV$ proton could travel through $5$ to $10$ layers and leave a short track, similar to those LDMX has proposed to use to reject short-lived charged kaon backgrounds~\cite{LDMX:2019gvz}. These higher-energy proton recoils may also be vetoed by NA64's selections on the lateral and longitudinal shape of the electromagnetic shower~\cite{Gninenko:2016kpg}. When the scattered nucleon is a neutron, it would miss the NA64 HCal completely due to the wide production angle and so is presumably undetectable, but the veto efficiency of the LDMX side HCal for wide-angle, low-energy neutrons is marginal (e.g.~see Fig.~50 of Ref.~\cite{LDMX:2018cma}). 

Properly determining the signal efficiency for incoherent meson production is thus an experimental question that requires detailed simulation and, preferably, in situ performance measurements. Nucleons with recoil energies below $50\ \MeV$ are virtually assured to appear as missing energy, while those with recoil energies up to $200\ \MeV$ would survive vetoes with an order one probability. For this work, we therefore take a kinetic energy cutoff of $100 \ \MeV$ for both protons and neutrons at NA64 and LDMX. However, as discussed below, our results are generally not qualitatively sensitive to the choice of cutoff.

We note that at LDMX, the $p_T$ distribution of missing-energy events from meson photoproduction and invisible decay matches that of ordinary Bremsstrahlung, not the higher-$p_T$ spectrum expected from $A'$ Bremsstrahlung. Therefore, electron $p_T$ does not offer any additional discriminating power between the meson-induced signal and Bremsstrahlung-initiated background. This does not impact our sensitivity analysis, which is based on LDMX projections that assume sub-single-event backgrounds \emph{before} any additional electron $p_T$ requirements. However, it does limit the toolkit available for distinguishing a meson-induced signal from mismodeled backgrounds. 

We will numerically estimate meson yields for four experimental benchmarks, described in Table~\ref{tab:yields}. First, we consider both existing NA64 data using a $100 \ \GeV$ beam, and the results of a future run with roughly $20$ times more electrons~\cite{Gninenko:2020hbd} at similar energies. At NA64, an event is potentially identified as signal if more than half of the energy is missing. On the basis of the thick target Bremsstrahlung results of Ref.~\cite{Tsai:1966js}, we estimate that a fraction $f_{\text{brem}} \approx 0.5$ of the electrons result in a hard photon carrying at least this much energy, and since the Bremsstrahlung spectrum is roughly flat, we take a typical photon energy of $75 \ \GeV$. For LDMX, we consider the two nominal stages of running described in Ref.~\cite{LDMX:2018cma}, assuming a thin, $10\%$ radiation length tungsten target. For the $4\ \GeV$ ``Phase I'' benchmark, the trigger requires the electron to lose more than $2.8 \ \GeV$ of its energy, which occurs via Bremsstrahlung to a fraction $f_{\text{brem}} \approx 0.03$ of the electrons, resulting in photons with typical energy $3.5 \ \GeV$. For the $8 \ \GeV$ ``Phase II'' benchmark, we double these energy numbers. 

\section{Vector Meson Photoproduction}
\label{sec:photoprod}

Our next task is to refine our estimate of the exclusive photoproduction yield $N_V$. Proton beam dumps face a similar problem, as in their case, pseudoscalar meson decay is an important source of DM. Typically, the reach of a proton beam dump experiment is estimated using Monte Carlo simulations (e.g.~see Refs.~\cite{MiniBooNE:2008hfu,Berlin:2018pwi,Dobrich:2019dxc}), which are tuned to match data at the $\sim 25\%$ level. However, this approach is unnecessary for our purposes. Particle transport Monte Carlo programs such as Geant4~\cite{Allison:2016lfl} excel at modeling the complex secondary interactions that occur for typical photons. This level of modeling is not required for our study, where photons only undergo a single, exclusive photoproduction process, and the vector mesons produced in these reactions decay well before interacting with any matter in the downstream detector.

In fact, a simulation-based approach is also inadequate, as neither Geant's hadronic models nor particle physics Monte Carlos such as Pythia~\cite{Sjostrand:2014zea} include careful modeling of exclusive photoproduction processes. Pythia's parton-based modeling is designed for the deep-inelastic regime, while our reactions of interest are in the diffractive regime. Meanwhile, Geant4~\cite{Wright:2015xia} does not include short-lived resonances in its hadronic models, but rather treats reactions such as $\gamma p \rightarrow \rho p$ as a component of, e.g., $\gamma p \rightarrow \pi^+\pi^- p$. More specialized programs such as GiBUU~\cite{Buss:2011mx} do propagate the light vector mesons through nuclei, but no semiclassical procedure can adequately describe coherent photoproduction, which often accounts for most of the meson yield. For heavy nuclei and high photon energies, neglecting coherent photoproduction underestimates the yield by up to an order of magnitude. 

Therefore, we will focus on estimating the yield transparently from a combination of theory and experimental measurements. Of course, to study more experimentally subtle questions, such as the probability of vetoing a recoiling nucleon from incoherent photoproduction, the models discussed below would need to be embedded into a Monte Carlo program with appropriate systematic uncertainties. 

To begin, we refine Eq.~\eqref{eq:prob_estimate} to account for the fact that the calorimeters are comprised of layers containing different nuclei. Weighting by the photon survival probability yields 
\begin{equation}
p_V = \int_0^\infty \exp\left( - \int_0^x \frac{7 \rho(x')}{9 X_0(x')} \, dx' \right) \frac{\rho(x) \sigma_0^V}{m_p} \, f_{\text{nuc}}^V(x) \, dx.
\end{equation}
For both NA64 and LDMX, this integral is largely determined by the composition of the front of the electromagnetic calorimeters. At NA64, most of the mesons are photoproduced in the lead absorber layers, with most of the remainder from carbon in the plastic scintillator~\cite{Bernhard:2020vca}. For LDMX, about half of the mesons are photoproduced from carbon, silicon, and oxygen in the preshower, while most of the rest are photoproduced in the tungsten absorber layers~\cite{LDMXEcal}. Note that in general, $p_V$ is higher for materials with lighter nuclei, because the radiation length scales as $X_0 \sim A/Z^2$, due to coherent scattering off the $Z$ protons.

\begin{figure}
\includegraphics[width=\columnwidth]{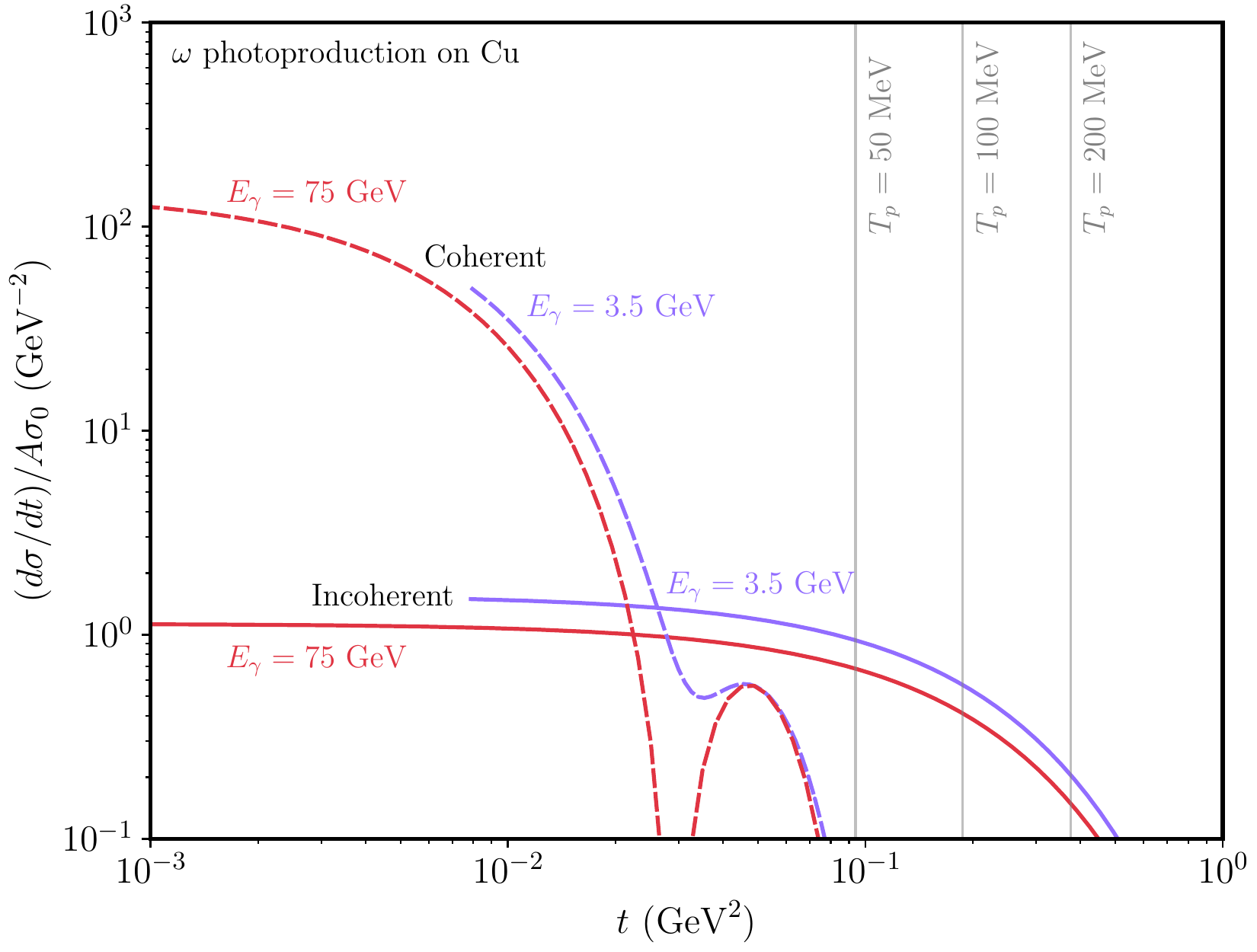}
\caption{Differential cross sections for coherent (dashed) and incoherent (solid) $\omega$ photoproduction. We also show contours of nucleon recoil energy for the incoherent process.}
\label{fig:differential_R}
\end{figure}

\begin{figure*}
\includegraphics[width=\columnwidth]{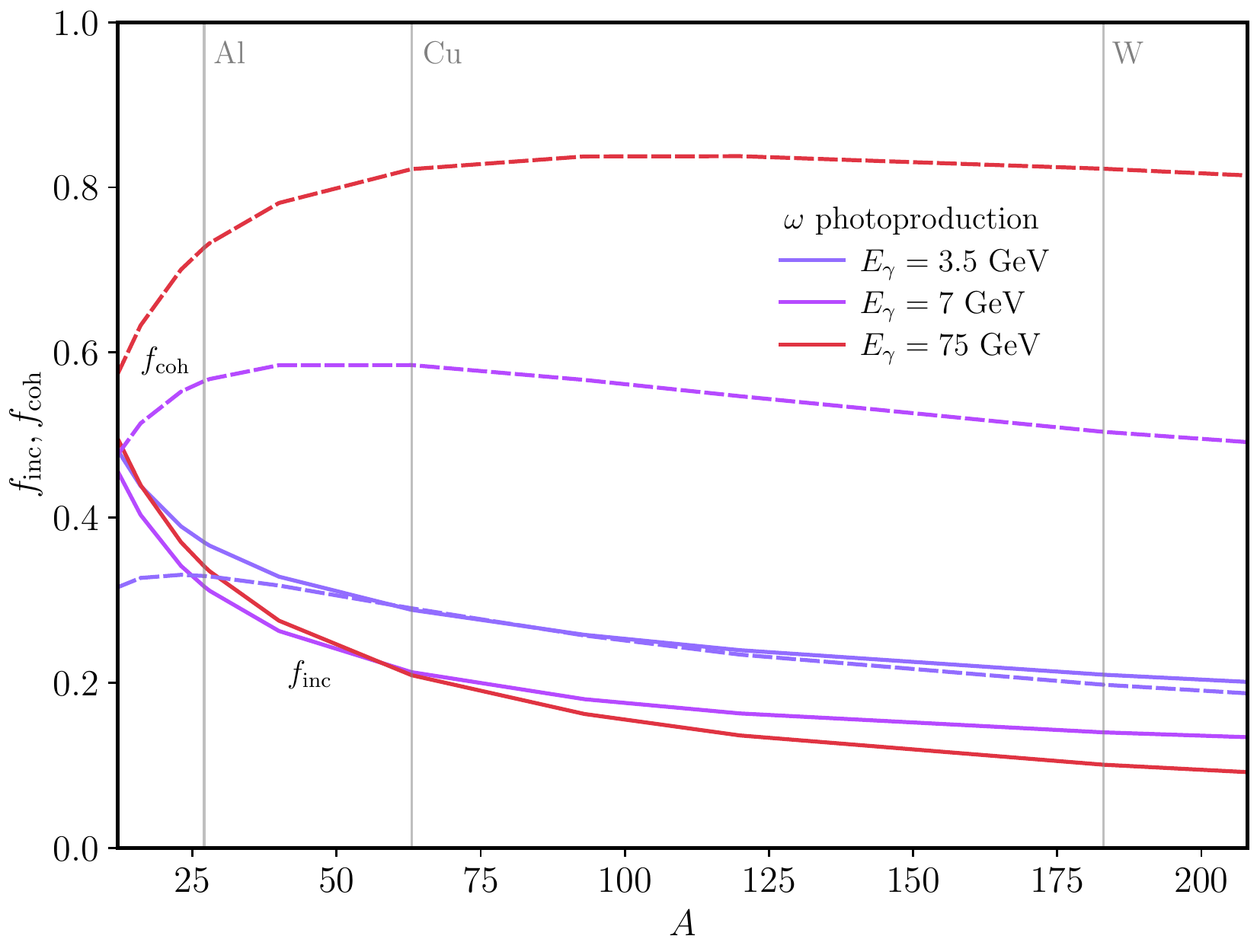}
\includegraphics[width=\columnwidth]{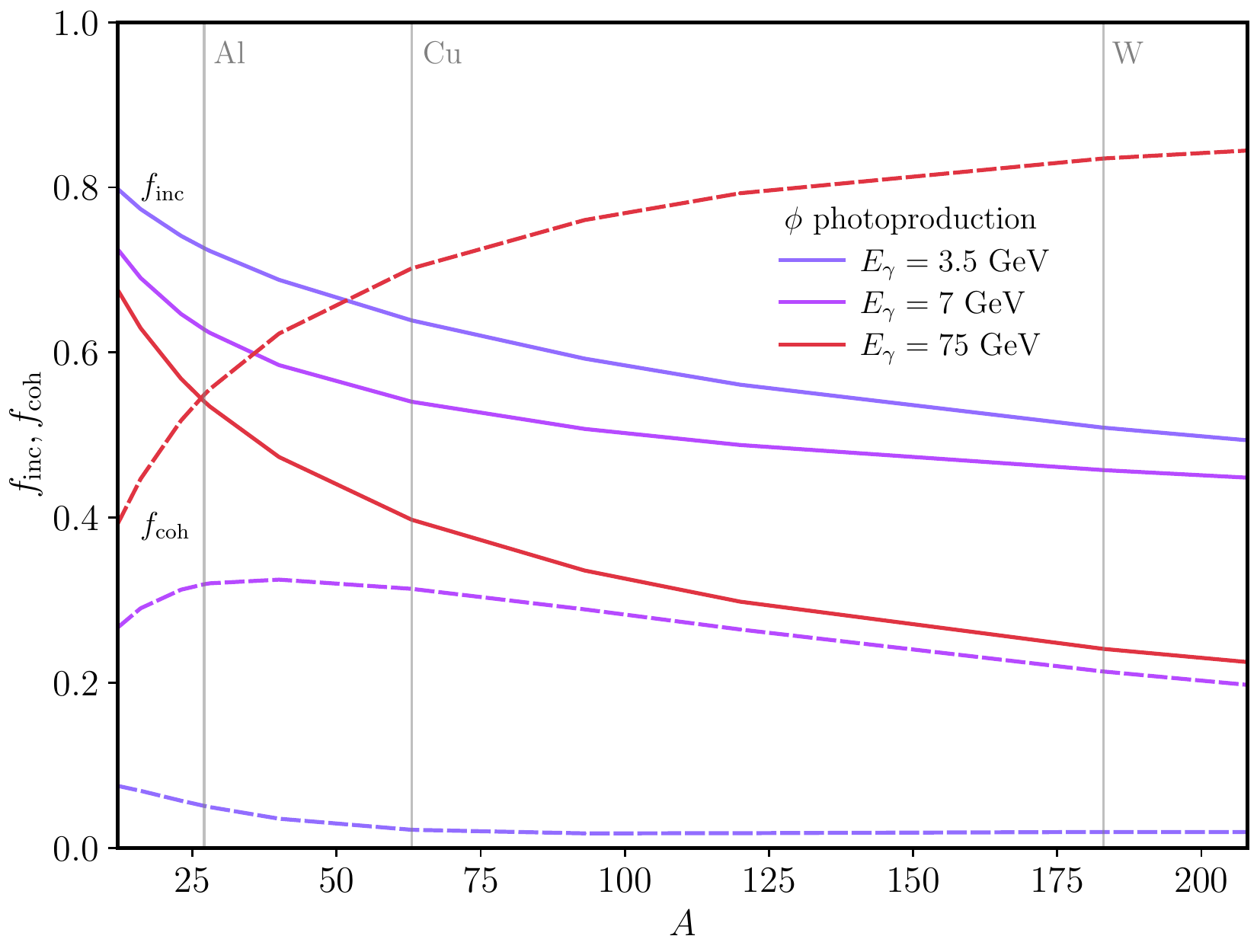}
\caption{Contributions to meson yield from incoherent (solid) and coherent (dashed) processes, as a function of the nucleon number $A$, for typical photon energies at LDMX and NA64. We show results for $\omega$ (left) and $\phi$ (right) photoproduction; results for $\rho$ are similar to those for $\omega$. Coherent production is suppressed for the $\phi$ meson at low photon energies, due to its higher mass.}
\label{fig:R_mass}
\end{figure*}

\begin{figure}
\includegraphics[width=\columnwidth]{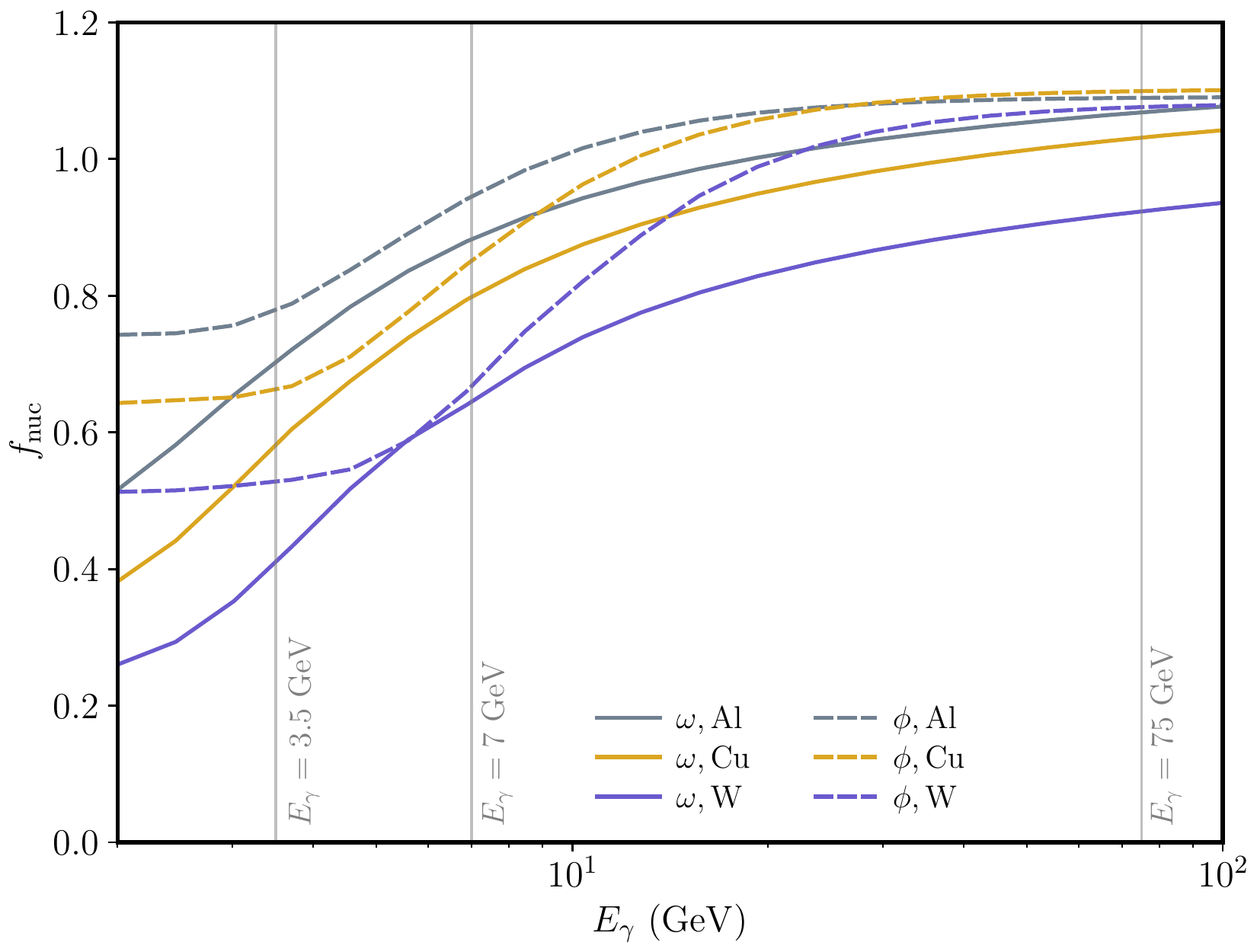}
\caption{Measure of total meson yield for the $\omega$ and $\phi$ for three benchmark nuclei, as a function of the photon energy $E_\gamma$.}
\label{fig:R_energy}
\end{figure}

\begin{table*}
\centering
\begin{tabular}{l|cccc|ccc|cccc}
& $N_e$ & $E_e$ & $E_\gamma$ & $f_{\text{brem}}$ & $\sigma_0^\rho \ (\mub)$ & $\sigma_0^\omega \ (\mub)$ & $\sigma_0^\phi \ (\mub)$ & $N_\rho$ & $N_\omega$ & $N_\phi$ & $N_{J/\psi}$ \\ \hline
NA64 (current) & $2.8 \times 10^{11}$ & $100 \ \GeV$ & $75 \ \GeV$ & $0.5$ & $9$ & $0.8$ & $0.7$ & $7 \times 10^6$ & $5 \times 10^5$ & $5 \times 10^5$ & $6 \times 10^3$ \\
NA64 (ultimate) & $5 \times 10^{12}$ & $100 \ \GeV$ & $75 \ \GeV$ & $0.5$ & $9$ & $0.8$ & $0.7$ & $1.2 \times 10^8$ & $9 \times 10^6$ & $8 \times 10^6$ & $1.1 \times 10^5$ \\
LDMX Phase I & $4 \times 10^{14}$ & $4 \ \GeV$ & $3.5 \ \GeV$ & $0.03$ & $23$ & $5$ & $0.4$ & $1.1 \times 10^9$ & $1.9 \times 10^8$ & $1.1 \times 10^7$ & -- \\
LDMX Phase II & $10^{16}$ & $8 \ \GeV$ & $7 \ \GeV$ & $0.03$ & $16$ & $1.9$ & $0.5$ & $3 \times 10^{10}$ & $3 \times 10^9$ & $5 \times 10^8$ & --\\
\end{tabular}
\caption{Total electrons on target $N_e$, electron energy $E_e$, and estimated fraction $f_{\text{brem}}$ of electrons that yield hard Bremsstrahlung photons (with typical energy $E_\gamma$), for our four benchmarks, as discussed in section~\ref{sec:expt}. We also show the per-nucleon exclusive light vector meson photoproduction cross sections and the estimated total meson yields, calculated as described in section~\ref{sec:photoprod}.}
\label{tab:yields}
\end{table*}

The per-nucleon exclusive photoproduction cross section $\sigma_0^V$ has been thoroughly measured for all relevant mesons in the entire energy range of interest. We extract the light vector meson cross sections in Table~\ref{tab:yields} from the theoretical fit of Ref.~\cite{Laget:2000gj}. For the $\phi$ meson, Pomeron exchange is the dominant contribution, explaining the characteristic slow rise in cross section with energy. For the $\rho$ and $\omega$ mesons, exchanges of light mesons such as the $f_2(1270)$ and $\pi^0$ dominate for low energies, while Pomeron exchange takes over at high energies, explaining why the cross section remains fairly high at NA64 energies. 

Naively, the cross section for nuclei is simply the incoherent sum $A \sigma_0^V$, but the computation of the correction factor $f_{\text{nuc}}$ is nontrivial. To proceed, we must note that mesons can be exclusively photoproduced through two distinct processes. In the coherent process, the nucleus remains in its ground state and recoils as a whole, leaving it with negligible kinetic energy; it is peaked at very low momentum transfer, with a scale set by the radius of the nucleus. In the incoherent process, the meson recoils off an individual nucleon, and the characteristic momentum transfers are somewhat higher, being set by Pomeron physics. These qualitative features are shown in Fig.~\ref{fig:differential_R}. 

We separate the contributions of these processes by defining $f_{\text{inc}, \text{coh}} = \sigma_{\text{inc}, \text{coh}} / A \sigma_0$, where $f_{\text{nuc}} = f_{\text{inc}} + f_{\text{coh}}$. Naively, the cross section for the incoherent process scales as $A$, but it is suppressed by absorptive final state interactions. In the limit of a very large, opaque nucleus, the effective number of nucleons participating is determined by the geometric cross section of the nucleus, $A_{\text{eff}} \sim r_{\text{nuc}}^2 \sim A^{2/3}$, leading to the rough scaling $f_{\text{inc}} \sim A^{-1/3}$. It is further suppressed by nuclear shadowing, a destructive interference effect most important at high photon energies. 

In the coherent process, the photoproduction amplitude is coherently summed over the nucleons, leading to a forward differential cross section $d \sigma_{\text{coh}} /dt |_{\theta = 0}$ that naively scales as $A^2$. The coherent peak extends up to $t \sim 1 / r_{\text{nuc}}^2 \sim A^{-2/3}$, implying a rough scaling $f_{\text{coh}} \sim A^{1/3}$, though it is also suppressed for heavy nuclei by absorptive final state interactions. In contrast to the incoherent process, the coherent cross section increases at higher photon energies, because a lower longitudinal momentum transfer $q_\parallel \approx m_V^2 / E_\gamma$ is required to produce the meson, leading to constructive interference across the entire nucleus. These qualitative features are shown in Fig.~\ref{fig:R_mass}.

The coherent and incoherent cross sections on nuclei can be measured separately, since coherent production is peaked at very low momentum transfer. For the light vector mesons, the coherent cross sections have been thoroughly measured decades ago, for a wide variety of nuclei and photon energies, and a standard Glauber optical model fits the data; we estimate that the theoretical uncertainty is at most 25\%. The subleading incoherent cross section is less well-measured, and the data is more ambiguous; here we estimate an uncertainty of up to 50\%. Further details on the theoretical modeling and experimental measurements may be found in the appendix, but for the purposes of estimating the reach, the conclusion is simply that $f_{\text{nuc}}$ is close to one for most energies and nuclei we consider, as shown in Fig.~\ref{fig:R_energy}. We compute the entries in Table~\ref{tab:yields} by additionally requiring that the nucleon recoil energy be less than $100\ \MeV$ for the incoherent process, which decreases the incoherent yield by up to 50\%. However, as shown in the appendix, our results are not strongly dependent on the precise choice of cutoff; it is usually the uncertainty on the coherent and incoherent cross sections that dominates. 

We may also consider heavier vector mesons, and the most promising example is $J/\psi$ photoproduction at NA64. Following HERA data~\cite{H1:2000kis}, we estimate a per-nucleon elastic photoproduction cross section of $\sigma_0^{J/\psi} = 15 \ \text{nb}$ at NA64 energies. Photoproduction of $J/\psi$ on nuclear targets may also be described by optical models, which have been recently refined within the leading twist approximation to treat ultraperipheral ion collisions at the LHC (e.g.~see Ref.~\cite{Guzey:2013jaa}). Fortunately, it is much more straightforward to treat the lower center-of-mass energies at NA64, $W_{\gamma p} \sim 10 \ \GeV$. In this case, the high longitudinal momentum transfer strongly suppresses coherent photoproduction and nuclear shadowing in incoherent photoproduction. Furthermore, final state interactions are relatively unimportant because the $J/\psi$-nucleon cross section is several times smaller than for the light vector mesons~\cite{Hufner:1997jg}. We may thus estimate $f_{\text{coh}} \approx 0$, $f_{\text{inc}} \approx 1$, with accuracy comparable to our other yield estimates.

NA64 can also produce $\Upsilon$ mesons, but the cross sections at its energies are orders of magnitude smaller than for $J/\psi$, making it uncompetitive with the current BaBar constraint. Meanwhile, at LDMX energies, $\Upsilon$ production and incoherent $J/\psi$ production are kinematically forbidden. Coherent $J/\psi$ production is kinematically allowed, but our optical models are not necessarily trustworthy in this very high momentum transfer regime; in any case, they predict a very strong suppression. We thus consider only $J/\psi$ production at NA64, yielding the final column of Table~\ref{tab:yields}.

\section{Invisible Branching Ratios}
\label{sec:ratios}

\begin{table}
\begin{tabular}{c|ccc}
$f_{V,X} \, (\MeV)$ & $\rho$ & $\omega$ & $\phi$ \\ \hline
$\bar{u} \gamma^\mu u$ & 157 & 136 & 8 \\
$\bar{d} \gamma^\mu d$ & $-148$ & 142 & 8 \\
$\bar{s} \gamma^\mu s$ & 0 & $-10$ & 233 \\ \hline
$J^\mu_{\text{EM}}$ & 154 & 46 & $-75$ \\
$J^\mu_B$ & 2.8 & 89 & 83 
\end{tabular}
\caption{Form factors describing the coupling of a light vector meson $V$ to a current $X$.}
\label{tab:quarks}
\end{table}

Given the meson yields shown in Table~\ref{tab:yields}, LDMX and NA64 can place a 90\% C.L.~limit $\Br(V \to \text{inv}) \leq 2.3 / N_V$ on the invisible branching ratio of each relevant meson, assuming zero background events and neglecting theoretical uncertainty, yielding the results shown in Fig.~\ref{fig:branching_summary}. We expect that current NA64 data could already set a strong limit on the invisible decays of the $\rho$ meson, improve the bounds for $\omega$ and $\phi$ by about an order of magnitude, and improve the $J/\psi$ bound by roughly a factor of $2$. Of course, actually setting such limits would require a more detailed analysis of theoretical uncertainties and experimental efficiencies, since our projections consider only statistical uncertainty.

Future NA64 data will improve on all of these results by roughly a factor of $20$. In particular, the resulting limit on $J/\psi$ invisible decay would be competitive with the projected limit $\Br(J/\psi \to \text{inv}) \leq 3 \times 10^{-5}$ from a future run of BES III~\cite{BESIII:2020nme}. LDMX could further improve on the light meson results by orders of magnitude, highlighting the potential of missing energy/momentum experiments as precision probes of meson physics. 

To constrain specific dark sector models, we must compute the expected invisible branching ratio of each meson. For concreteness, we focus on dark sectors with pseudo-Dirac fermion DM, and a vector mediator with interactions 
\begin{equation}
\mathcal{L} \supset \epsilon e A'_\mu J^\mu_{X} + g_D \, A'_\mu \bar{\chi} \gamma^\mu \chi
\end{equation}
where $\epsilon$ is the kinetic mixing parameter, and $J^\mu_{X} = \sum_q c_q \bar{q} \gamma^\mu q$ stands for the electromagnetic current $J^\mu_{\text{EM}}$ ($c_q = Q_q$) or the baryon current $J^\mu_B$ ($c_q = 1/3$). For the $U(1)_B$ model, we also assume a loop-suppressed kinetic mixing $\epsilon e / (4 \pi)^2$ with the photon. This does not affect the meson decay signature, but is relevant for computing the reach from $A'$ Bremsstrahlung and the thermal relic target below the two-pion threshold. For the dark sector parameters, we take benchmarks $\alpha_D = g_D^2 / 4 \pi = 0.5$ and $m_{A'} / m_\chi = 3$. 

To compute the invisible branching ratios for light vector mesons, we first find the relevant form factors, defined by $\langle 0 | J^\mu_{X} | V(\epsilon) \rangle = i m_V f_{V,X} \epsilon^\mu$. From the results of Appendix C of Ref.~\cite{Bharucha:2015bzk}, which accounts for $\phi$-$\omega$ and $\rho$-$\omega$ mixing, we infer the first three rows of Table~\ref{tab:quarks}, giving the couplings $f_{V,q}$ to light quark currents $\bar{q} \gamma^\mu q$. Note that $f_{\rho, s}$ vanishes because we are ignoring the small effect of $\rho$-$\phi$ mixing, but this does not qualitatively affect the results. Next, we straightforwardly infer the bottom two rows of Table~\ref{tab:quarks}. Here, $f_{\omega, \text{EM}}$ is suppressed due to a partial cancellation between the $u$ and $d$ components, while $f_{\rho, B}$ is strongly suppressed, because it is only nonzero due to mixing effects. 

The amplitude for an on-shell vector meson $V$ to decay to $\chi \bar{\chi}$ through a virtual $A'$ is 
\begin{equation}
\mathcal{M} = \frac{\eta^{\mu\nu} - q^\mu q^\nu / m_{A'}^2}{q^2 - m_{A'}^2} \, (g_D \epsilon e) (\bar{u}(p) \gamma_\nu v(p')) (m_V f_{V,X} \epsilon_\mu)
\end{equation}
where $q^\mu$ is the meson's momentum. Squaring and summing over final spins and averaging over initial meson polarizations $\epsilon^\mu$ gives the decay rate 
\begin{equation} \label{eq:differential_gamma}
\Gamma_{V \to \chi \bar{\chi}} = \frac{\alpha_D (\epsilon e)^2 f_{V,X}^2}{3} \frac{(m_V^2 + 2 m_\chi^2)\sqrt{m_V^2 - 4 m_\chi^2}}{(m_{A'}^2 - m_V^2)^2 + \Gamma_{A'}^2 m_{A'}^2}
\end{equation}
in the frame of the meson. The resonant peak is cut off by the width $\Gamma_{A'}$ of the $A'$ due to decay to DM, where for our dark sector model, 
\begin{equation}
\frac{\Gamma_{A'}}{m_{A'}} = \frac{\alpha_D}{3} \sqrt{1 - 4 m_\chi^2/m_{A'}^2} \ (1 + 2 m_\chi^2/m_{A'}^2).
\end{equation}
for on-shell $A'$. (These results can equivalently be derived by considering mixing with the $A'$ in the vector meson dominance framework, e.g.~see Ref.~\cite{Batell:2014yra}.) The expected number of missing energy events via production of $V$ is then $N_V \Gamma_{V \to \chi \bar{\chi}} / \Gamma_V$.

However, in the above derivation, we have implicitly applied the narrow width approximation for the vector meson $V$ by taking it to be on-shell. A more accurate expression is obtained by averaging Eq.~\eqref{eq:differential_gamma} over the spectral density of photoproduced $V$'s, or equivalently, by treating both $V$ and $A'$ as intermediate states in the full $\gamma N \to \chi \bar{\chi} N$ process. In particular, when $\Gamma_{A'} \ll \Gamma_V$ but the two resonances overlap, $|m_{A'} - m_V| \lesssim \Gamma_V$, it is more accurate to apply the narrow width approximation to the $A'$. Assuming a Breit--Wigner lineshape for both resonances, the resulting correction is equivalent to multiplying Eq.~\eqref{eq:differential_gamma} by $\Gamma_V / \Gamma_{A'}$ and replacing $\Gamma_{A'}$ with $\Gamma_V$ in the denominator. We apply this correction when showing results for the $\rho$ meson, which is about twice as wide as the $A'$ for nearby masses, resulting in a flattening and broadening of the resonant peaks.

When the masses are widely separated, $|m_{A'} - m_V| \gg \Gamma_{A'}, \Gamma_V$, the situation is more subtle. Here, the spectral density has two distinct contributions. The contribution at $q^2 \approx m_V^2$ corresponds to the standard result from on-shell $V$'s, while the additional contribution at $q^2 \approx m_{A'}^2$ corresponds to production of far off-shell vector mesons that mix with a nearly on-shell $A'$. The contribution of this second peak can be comparable or even greater, especially when $\alpha_D \ll 1$. It cannot be interpreted in terms of an invisible branching ratio of the $\rho$, but it does enhance the signal rate. However, properly evaluating this contribution would require a more detailed treatment of the momentum-dependence of the photon-$\rho$ Pomeron vertex, the final-state phase space, and the spectral shape of the $\rho$. Therefore, we conservatively neglect it in this initial study.

Finally, for the $J/\psi(1S)$, we compute the invisible decay width by comparing it to the decay width to $e^+ e^-$, as in Ref.~\cite{Wilczek:1977zn}. Because the quarks carry spin $1$, the quark spinor bilinear $\bar{u} \gamma^\mu v$ is purely spatial, which implies that the longitudinal term in the $A'$ propagator does not contribute to the amplitude. As a result, the two decay widths are identical up to constants and kinematic factors, giving
\begin{multline}
\frac{\Br(J/\psi \to \chi \bar{\chi})}{\Br(J/\psi \to e^+ e^-)} = \frac{\alpha_D}{\alpha_e} \left(\frac{c_c \epsilon}{Q_c} \right)^2 \\ \times \frac{m_V (m_V^2 + 2 m_\chi^2)\sqrt{m_V^2 - 4 m_\chi^2}}{(m_{A'}^2 - m_V^2)^2 + \Gamma_{A'}^2m_{A'}^2} 
\end{multline}
where we used $m_e \ll m_V = m_{J/\psi}$. In the limit $m_\chi, m_{A'} \ll m_V$, the final factor reduces to unity, leaving the expected ratio of couplings.

\begin{figure*}
\includegraphics[width=\columnwidth]{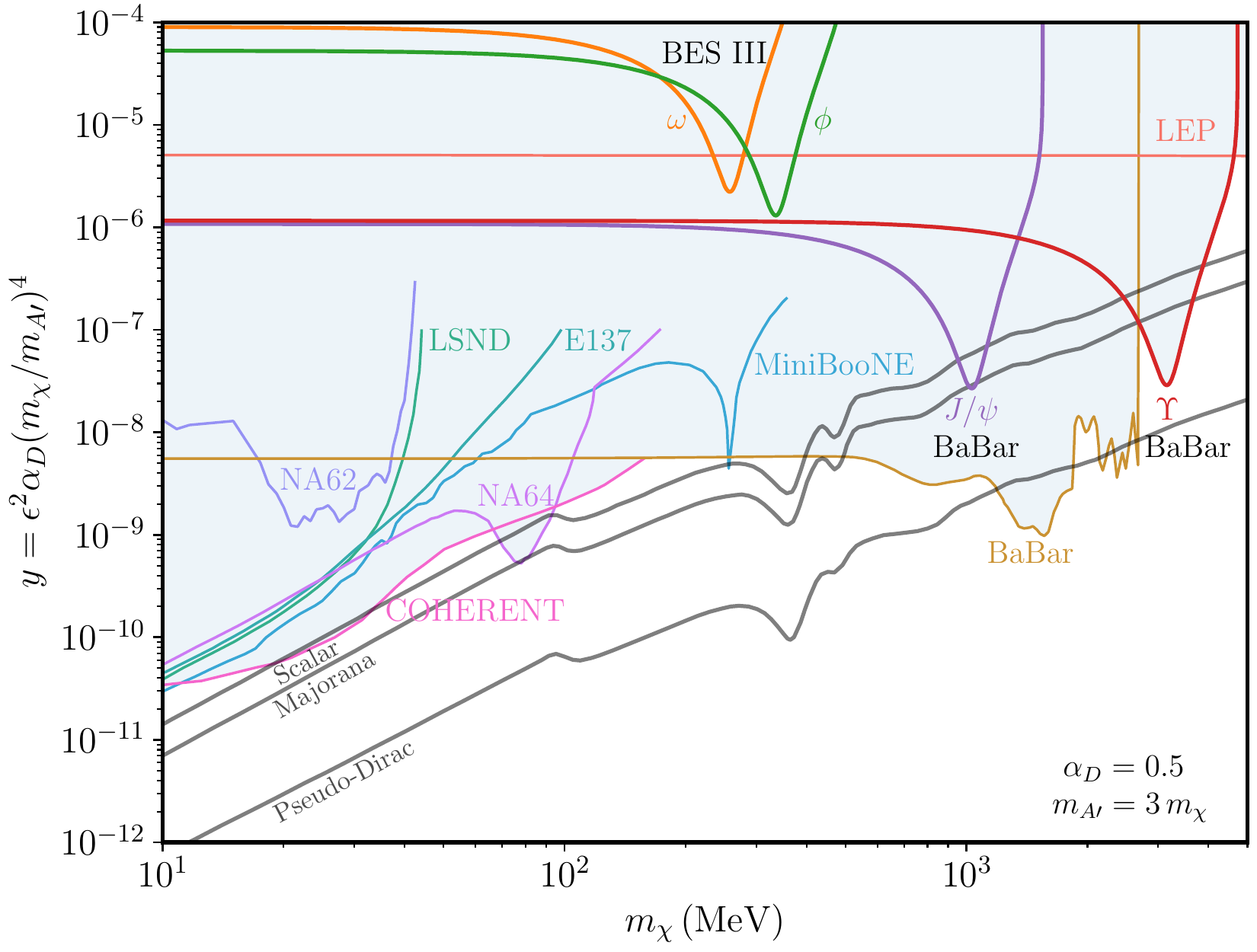}
\includegraphics[width=\columnwidth]{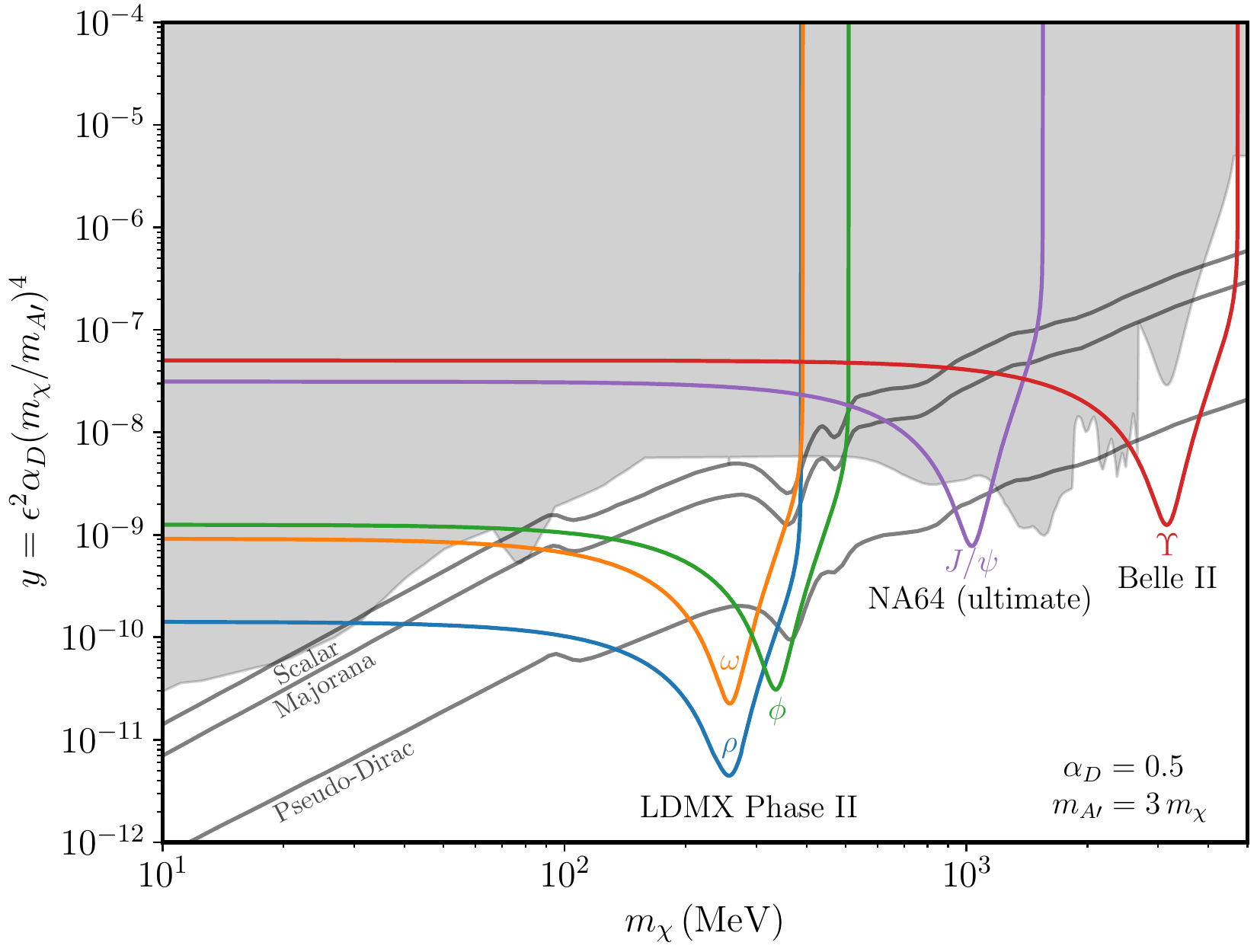}
\caption{Constraints on the dark photon model with fermionic DM. At left, we show existing constraints from current bounds on invisible vector meson decays, as well as those from the beam dumps LSND~\cite{LSND:2001akn,deNiverville:2011it}, E137~\cite{Bjorken:1988as,Batell:2014mga}, MiniBooNE~\cite{MiniBooNEDM:2018cxm}, and COHERENT~\cite{Akimov:2021yeu}, radiative pion decay at NA62~\cite{NA62:2019meo}, the missing energy experiment NA64~\cite{Banerjee:2019pds,Andreev:2021fzd}, production in $e^+ e^-$ collisions at BaBar~\cite{BaBar:2017tiz}, and precision measurements of the $Z^0$ mass at LEP~\cite{Hook:2010tw}. At right, we show the strongest projected 90\% C.L.~exclusions from invisible decays of each meson alone. For the light vector mesons and $J/\psi$, these constraints will come from LDMX and NA64, respectively, assuming zero background events. For the $\Upsilon$, we take the projected limit $\Br(\Upsilon \to \text{inv}) < 1.3 \times 10^{-5}$ from Belle II~\cite{Belle-II:2018jsg}.}
\label{fig:dark_photon_constraints}
\end{figure*}

\begin{figure*}
\includegraphics[width=\columnwidth]{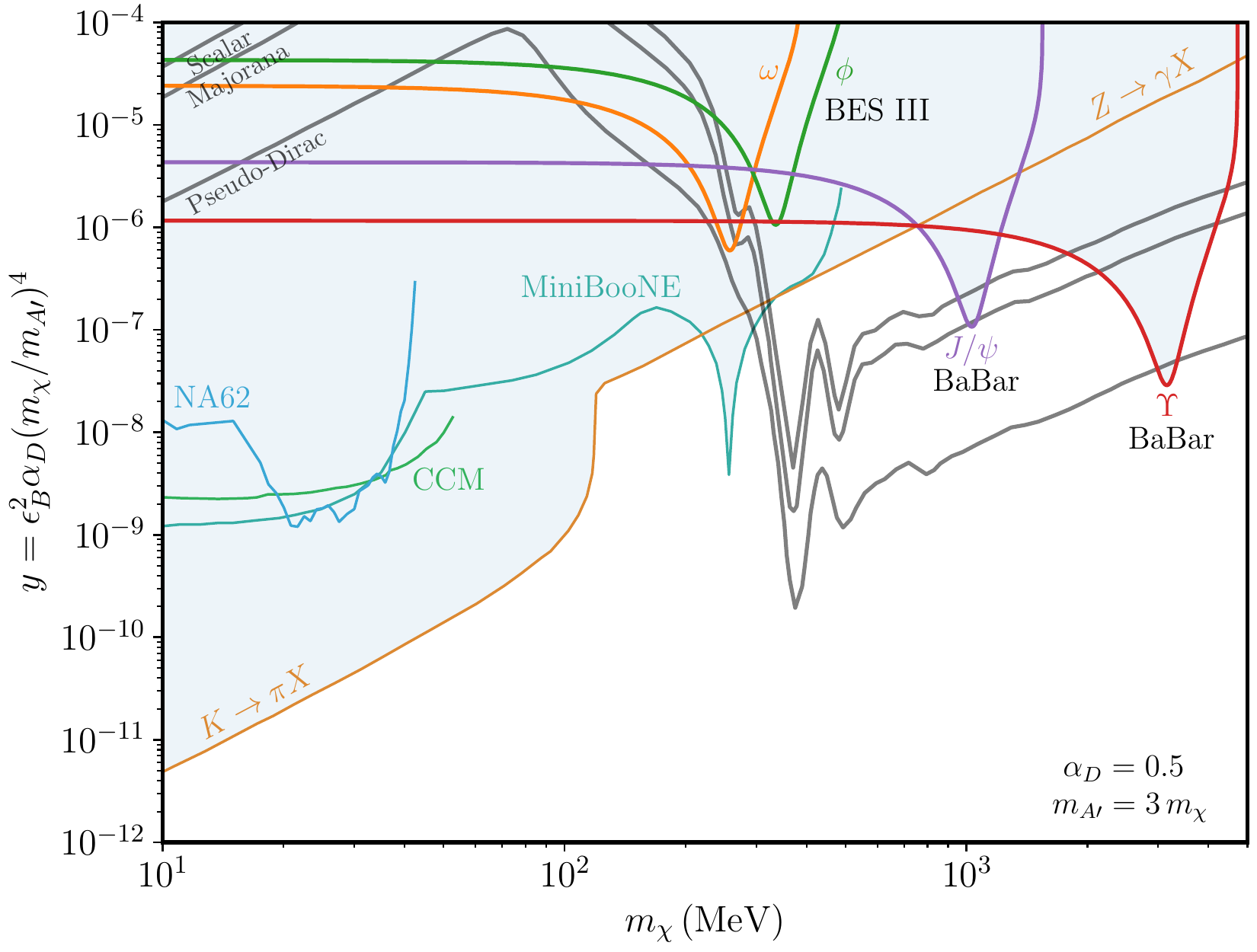}
\includegraphics[width=\columnwidth]{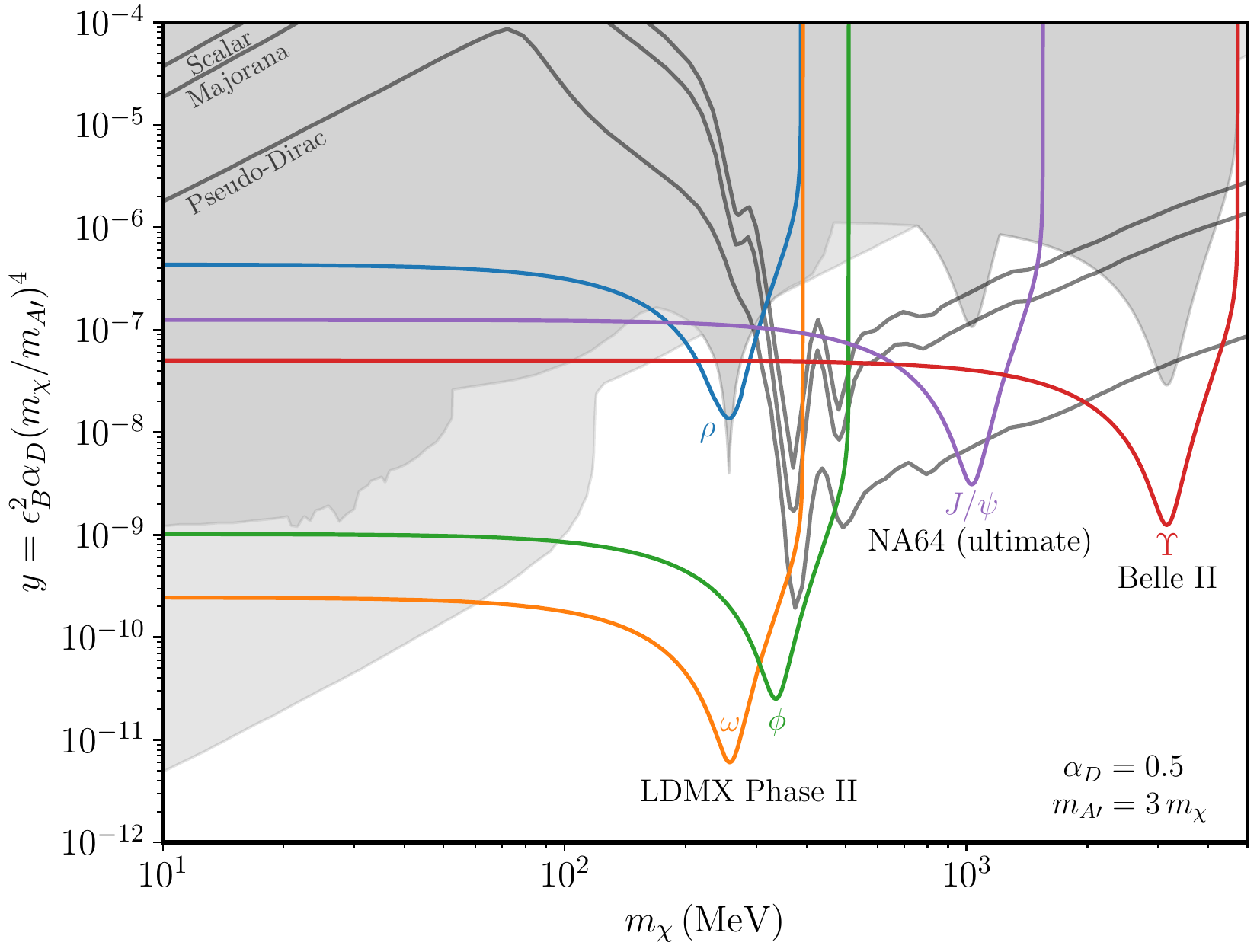}
\caption{Same as Fig.~\ref{fig:dark_photon_constraints}, but in the $U(1)_B$ model. In this case, the leading existing constraints are from CCM~\cite{CCM:2021leg,Aguilar-Arevalo:2021sbh} (rescaled to $m_{A'} / m_\chi = 3$ using Ref.~\cite{Berlin:2020uwy}), NA62~\cite{NA62:2019meo} (recast for a $U(1)_B$ gauge boson using Ref.~\cite{Batell:2014yra}), MiniBooNE~\cite{MiniBooNEDM:2018cxm} (rescaled to $\alpha_D = 0.5$), and rare processes $K \to \pi X$ and $Z \to \gamma X$, which have $1/m_{A'}^2$ enhanced rates due to the $U(1)_B$ anomaly~\cite{Dror:2017ehi,Dror:2017nsg}. These latter constraints are shaded more lightly at right, since they may be removed by coupling to a nonanomalous current such as $B - 3 L_\tau$. Thermal relic curves assume a loop-suppressed kinetic mixing with the photon, and are computed as in Ref.~\cite{Berlin:2018bsc}.}
\label{fig:baryon_constraints}
\end{figure*}

\begin{figure*}
\includegraphics[width=\columnwidth]{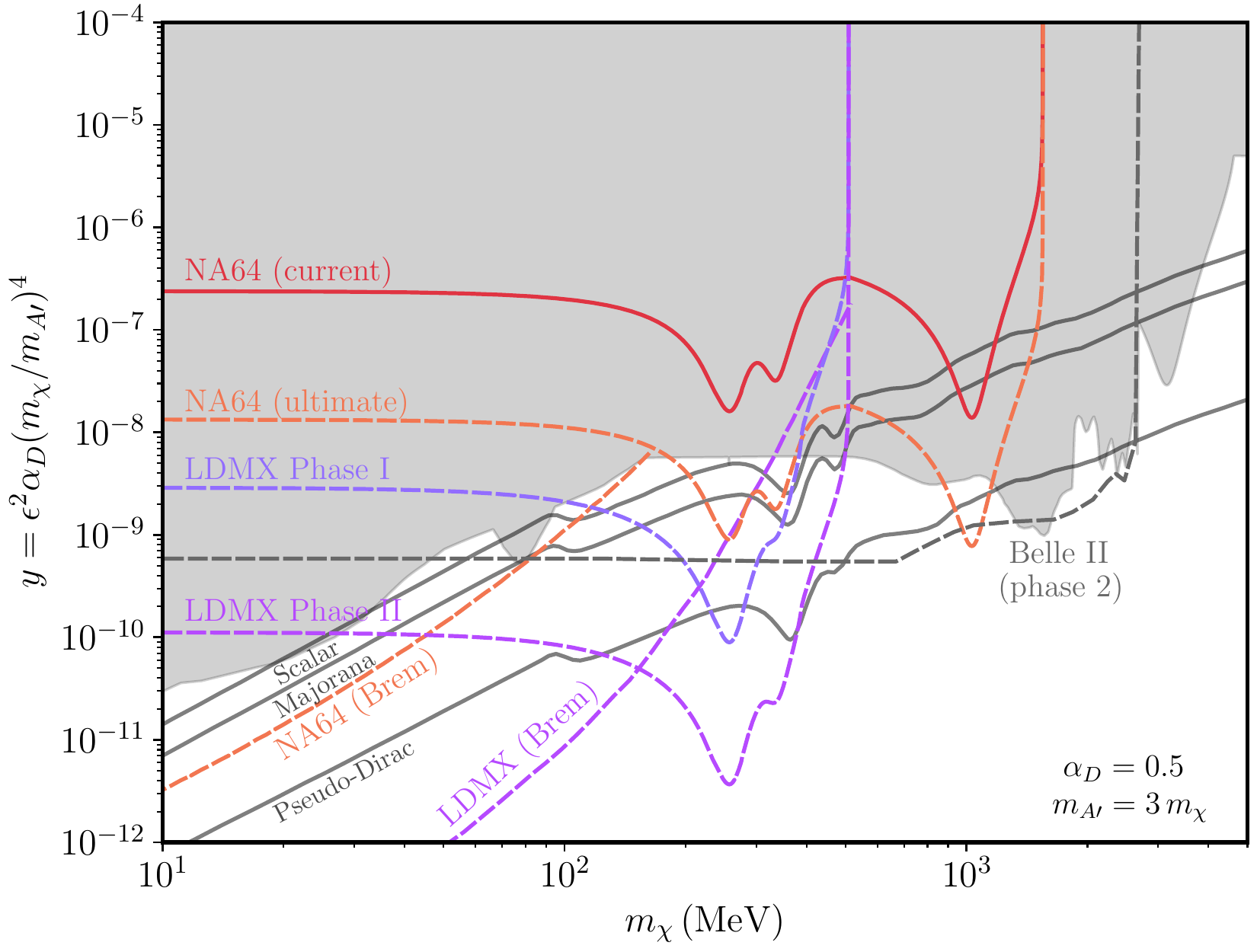}
\includegraphics[width=\columnwidth]{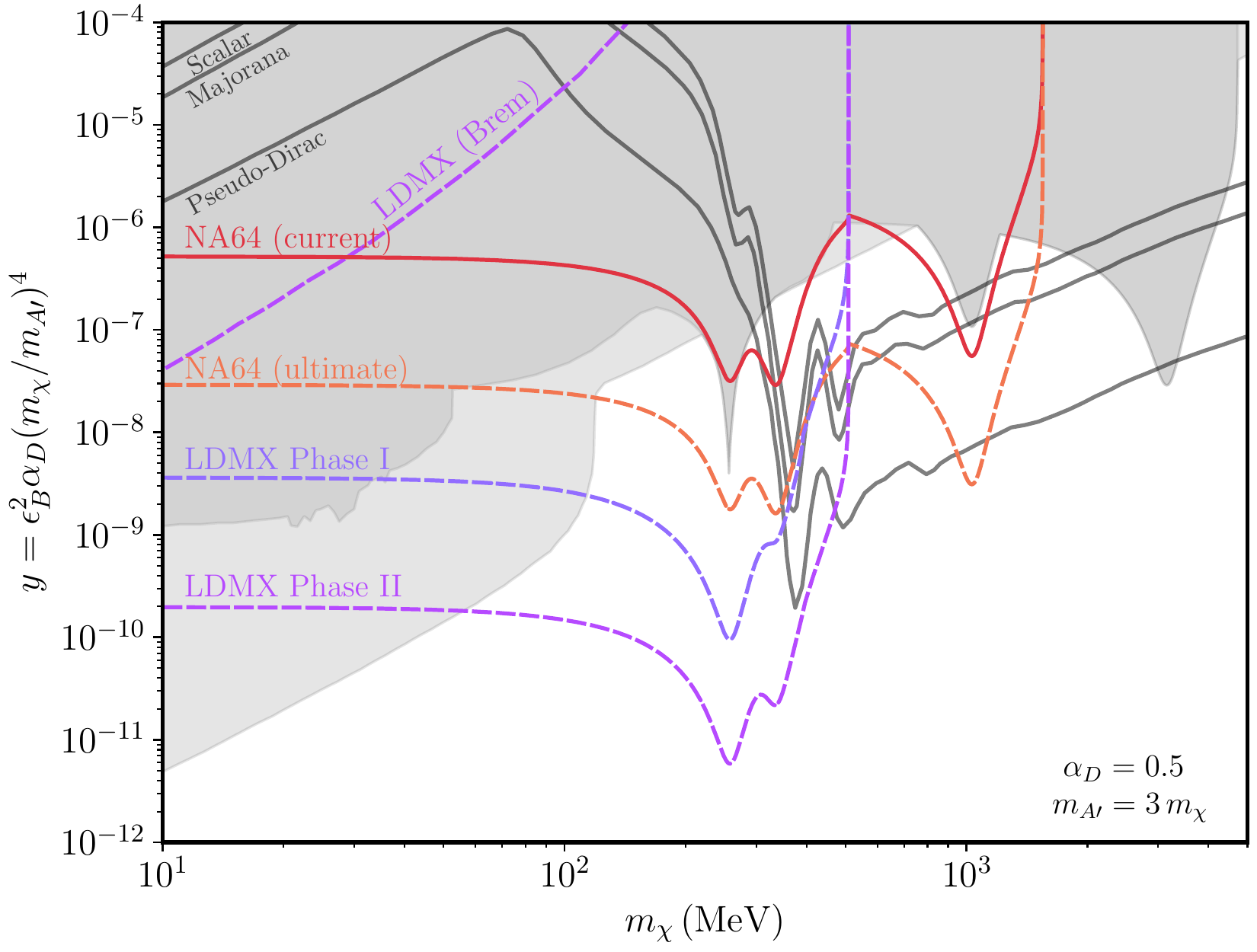}
\caption{Projected 90\% C.L.~exclusions for the four experimental benchmarks described in Table~\ref{tab:yields}, assuming zero background events, in the dark photon (left) and $U(1)_B$ (right) models, both with fermionic DM. For comparison, we also show projections for Belle II~\cite{Belle-II:2018jsg}, and for $A'$ Bremsstrahlung at LDMX Phase II~\cite{Berlin:2018bsc} and a future run of NA64~\cite{Gninenko:2020hbd}. For projections from proton beam experiments, see Ref.~\cite{deNiverville:2016rqh}.}
\label{fig:expt_constraints}
\end{figure*}

\section{Projected Reach}
\label{sec:results}

For a specific dark sector model, the expected number of signal events is $N_{\text{sig}} = \sum_V N_V \Br(V \to \text{inv})$, and if no signal and background events are seen, a 90\% C.L.~limit may be set on the model by imposing $N_{\text{sig}} < 2.3$. In Fig.~\ref{fig:dark_photon_constraints}, we show the current and projected constraints on the dark photon model from individual mesons. As indicated in the left panel, current constraints on the invisible decays of $J/\psi$ and the light vector mesons are not competitive with the leading experimental bounds. However, as shown in the right panel, missing energy experiments can improve the latter constraints by up to $5$ orders of magnitude, in the case of LDMX Phase II. When combined with the projected improved measurements of $J/\psi$ and $\Upsilon$ invisible decays, we find that meson decays alone can probe a substantial portion of the thermal freeze-out ``target'' region of couplings for $m_\chi \gtrsim 0.1 \ \GeV$. 

In Fig.~\ref{fig:baryon_constraints}, we show the same results for a $U(1)_B$ gauge boson mediator. In this case, most constraints on dark photons do not apply, because they depend on the dark photon's coupling to electrons. Instead the strongest constraints for most masses come from precision measurements of rare processes, which are enhanced due to the $U(1)_B$ anomaly, though invisible $\Upsilon$ decay remains the strongest constraint at high masses. Despite these strong existing bounds, the projected future reach from meson decay probes new parameter space for $m_\chi \gtrsim 0.1 \ \GeV$, and covers almost the entirety of the thermal freeze-out region. The sensitivity to invisible $\rho$ meson decay is suppressed by a small form factor, as noted above, but this is compensated by increased sensitivity to invisible $\omega$ decays. 

Referring to Table~\ref{tab:branching_rates}, invisible decays to neutrinos in the Standard Model remain a negligible background for the light vector mesons, even at LDMX Phase II. However, they imply that any future meson decay experiment will only be able to probe $2$ to $3$ orders of magnitude beyond LDMX Phase II, before running into a ``neutrino floor'' that slows further progress. For the $J/\psi$, this floor is a few orders of magnitude beyond the NA64 (ultimate) projections, while Belle II will nearly reach the $\Upsilon$ floor. Conversely, by searching for invisible meson decays, LDMX and NA64 will also be sensitive to neutrino-quark interactions from physics beyond the Standard Model. As we discuss in section~\ref{subsec:neutrino}, they could provide the leading bounds on several effective operators that are otherwise difficult to probe with upcoming neutrino experiments. 

In Fig.~\ref{fig:expt_constraints}, we show the constraints on dark sectors for our four experimental benchmarks. The main qualitative feature is that in all cases, meson decay improves the reach of these experiments for $m_{A'} \gtrsim 0.5 \ \GeV$. In this regime, the reach due to $A'$ Bremsstrahlung is weak, due to the $(m_e/m_{A'})^2$ suppression of the Bremsstrahlung cross section, and the fact that the higher momentum transfer begins to resolve heavy nuclei. In the case of NA64, meson decay does not currently probe new parameter space. However, it will allow future runs of NA64 to probe the thermal freeze-out region at masses around the resonances $m_{A'} \approx m_V$, up to an order of magnitude higher in mass than through $A'$ Bremsstrahlung alone. Meson decay also extends the reach of LDMX upward in mass, by roughly a factor of $2$. We note that the reach from Belle II is highly complementary: when combined with LDMX, the thermal freeze-out region from $\MeV$ to $\GeV$ masses will be well explored.

For the $U(1)_B$ model, meson decay dramatically improves the reach of both NA64 and LDMX, since the $A'$ Bremsstrahlung channel is penalized by the loop-suppressed coupling to electrons. Typically, these experiments are viewed as probing mediator couplings to electrons, while proton beam experiments probe couplings to quarks. The meson decay signature discussed here shows that electron beam experiments can have competitive sensitivity to quark couplings. 

The constraints we show fall sharply at threshold, $2 m_\chi = m_V$, but in fact, heavier dark matter can be produced in these experiments through the decays of heavier mesons, such as the resonances $\omega(1420)$ and $\rho(1450)$ of the light vector mesons. The main obstacle to predicting this sensitivity is the lack of data on photoproduction cross sections for these resonances. We expect the sensitivity to be lower, due to the larger width of the resonances, but potentially still high enough to probe new parameter space. Similarly, for the $J/\psi$, one can consider production of excited charmonium states such as $\psi(2S)$. 

It is also interesting to consider how the reach due to meson decay depends on the dark sector parameters $\epsilon$, $\alpha_D$, $m_\chi$, and $m_{A'}$. First, both the thermal annihilation cross section and the invisible meson decay rate are proportional to $\epsilon^2 \alpha_D$, which implies that for a fixed mass ratio $R = m_{A'} / m_\chi$, both of these curves are roughly independent of $\alpha_D$ on a $y$ vs.~$m_\chi$ plot. Therefore, invisible meson decay maintains its relative sensitivity to the thermal freeze-out region for lower $\alpha_D$, though the reach due to $A'$ Bremsstrahlung improves. (However, as mentioned in Section~\ref{sec:ratios}, the invisible decay rate might actually be enhanced at lower $\alpha_D$ because of the contribution of off-shell vector mesons mixing with on-shell $A'$.)

\begin{figure}
\includegraphics[width=\columnwidth]{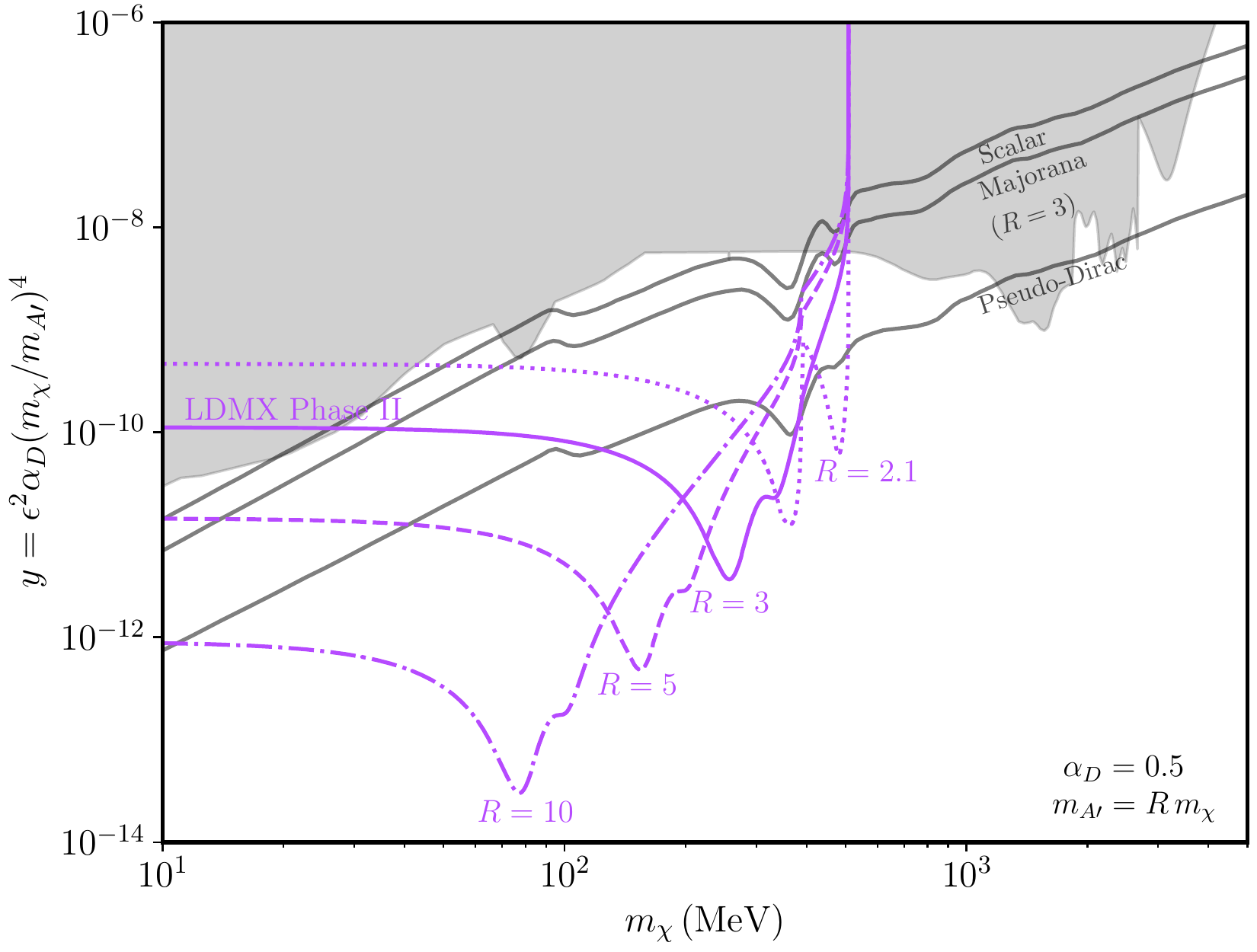}
\caption{Projected 90\% C.L.~exclusions from invisible vector meson decay at LDMX Phase II in the dark photon model, for several choices of the mass ratio $R = m_{A'} / m_\chi$. Thermal targets are shown for $R = 3$, and are roughly independent of the mass ratio except when $R-2$ is small.}
\label{fig:param_variation}
\end{figure}

The effect of changing the mass ratio $R$ is shown in Fig.~\ref{fig:param_variation}. For low $m_{A'}$, increasing $R$ rapidly improves the reach in $y$ of both $A'$ Bremsstrahlung and meson decay, because $y \propto 1/R^4$, while in this regime the $A'$ Bremsstrahlung rate is proportional to $1/R^2$, and the invisible meson decay rate is independent of $R$. Increasing $R$ allows invisible meson decay to probe very high mediator masses, $m_{A'} \gg m_V$, even though the mixing with the vector meson is lowered. In this regime $A'$ Bremsstrahlung would be strongly suppressed because of the high momentum transfer required to produce the $A'$, but invisible meson decay is not, as the momentum transfer is set by $m_V$ rather than $m_{A'}$. 

Because invisible meson decay proceeds through off-shell mediators, it also occurs in the ``forbidden'' regime $R < 2$, where an on-shell $A'$ cannot decay to DM. In this case, there may be complementary constraints from visible $A'$ decay, and there is still a predictive thermal target for sufficiently high $m_{A'}$ and $\alpha_D$~\cite{Berlin:2018bsc}. However, the potential to reach these targets through the meson decay signature is weaker because of the scaling with $R$ mentioned in the previous paragraph. 

\section{Future Directions}
\label{sec:disc}

\subsection{Additional Meson Decays}
\label{subsec:add_meson}

We have focused on the invisible decay of vector mesons as a probe of dark sectors with vector mediators, but there are numerous potential extensions. First, missing energy experiments will also set leading, but weaker limits on the invisible decay of pseudoscalar mesons. NA62 has set a strong bound on invisible $\pi^0$ decay, which seems difficult to improve upon, but the bounds on invisible $\eta$ and $\eta'$ decay are quite weak. Since these mesons have a different spin from the photon, they cannot be photoproduced by Pomeron exchange. Instead, their exclusive photoproduction is described by Reggeon exchange, leading to cross sections that rapidly fall with energy~\cite{Mathieu:2015eia,JPAC:2016lnm}. This suppresses the yield of these mesons at NA64, though both NA64 and LDMX should still be able to significantly improve bounds for $\eta$ and $\eta'$. 

In a similar vein, it could be interesting to investigate the invisible decay of scalar mesons ($J^{PC} = 0^{++}$), such as the $f_0(980)$ and $a_0(980)$, as these provide complementary information to the invisible decay of vector mesons. However, there is currently little data on the photoproduction of these mesons, with the first measurements only made comparatively recently~\cite{CB-ELSA:2008zkd,CLAS:2008ycy}.

The invisible decay of neutral kaons has never been measured, and both NA64 and LDMX should be able to set strong bounds, without requiring a kaon beam. We note, however, that there are several complications that suppress the potential sensitivity. First, as for $\eta$ and $\eta'$, the exclusive photoproduction cross sections for $K_S$ and $K_L$ fall rapidly with energy. Due to conservation of strangeness, kaon photoproduction converts a nucleon to a hyperon. This implies a minimum energy deposition in the calorimeters of about $200 \ \MeV$ when the hyperon decays; determining the associated veto efficiency requires further detector study. Finally, the $K_L$ is sufficiently long-lived that there is a substantial penalty from demanding that it decay before interacting with the calorimeters.

Mesons can also undergo radiative decays, into a photon plus missing energy. For both LDMX and NA64, the signal efficiency for such a process is penalized because the photon must carry sufficiently little energy. This is relatively unimportant at NA64, where the photon may carry a substantial fraction of the meson's energy, but at LDMX we require the photon to be soft enough to completely avoid detection in the calorimeters, which requires it to be emitted nearly backwards in the meson's frame. It may be possible to partially circumvent this penalty by allowing a small amount of energy to be detected in the calorimeters, but this could introduce other backgrounds, which would require a more detailed study to evaluate.

In particular, missing energy experiments could improve bounds on radiative $\eta$ and $\eta'$ decay by orders of magnitude, extending existing constraints associated with radiative $\pi^0$ decay to higher $A'$ mass. Constraints on this process could also be set using existing data at BES~\cite{Fang:2017qgz} and KLOE~\cite{Kang:2019eby}, and much stronger future constraints could come from ``$\eta$-factory'' experiments~\cite{Gan:2020aco}, such as the upcoming JEF~\cite{JEF} and proposed REDTOP~\cite{Gatto:2019dhj}.

\subsection{Additional Dark Sectors}

\begin{table}
\centering
\begin{tabular}{c|cccc}
mediator coupling& $V \to \chi \bar{\chi}$ & $V \to \gamma \chi \bar{\chi}$ & $M \to \chi \bar{\chi}$ & $M \to \gamma \chi \bar{\chi}$ \\ \hline
scalar $\bar{q} q$ & & $\checkmark$ & & \\
pseudoscalar $\bar{q} \gamma^5 q$ & & $\checkmark$ & $\checkmark$ & \\
vector $\bar{q} \gamma^\mu q$ & $\checkmark$ & & & $\checkmark$ \\
axial vector $\bar{q} \gamma^\mu \gamma^5 q$ & & $\checkmark$ & $\checkmark$ & \\
\end{tabular}
\caption{Quark couplings to mediators that can cause invisible and radiative decays of vector mesons $V$ ($J^{PC} = 1^{--}$) and pseudoscalar mesons $M$ ($J^{PC} = 0^{-+}$).}
\label{tab:allowed}
\end{table}

We have focused on dark sectors with vector mediators because they are probed by invisible vector meson decay, but as indicated in Table~\ref{tab:allowed}, the other meson decays discussed in the previous section can probe other types of mediators. Here we consider each possibility in turn. 

First, we note that vector mediators can also be probed through radiative pseudoscalar meson decays. For the reasons given in the previous subsection, we expect that the associated constraints will be weaker, for the models considered in this paper, than those from invisible vector meson decays. On the other hand, when $m_{A'} < m_M$, the $A'$ can be produced on-shell in the decay $M \to A' \gamma$, giving an radiative decay rate independent of $\alpha_D$. It could therefore be a leading constraint in models with $\alpha_D \ll 1$, such as $U(1)_B$ models with $\alpha_D \sim \alpha_B$. 

Scalar mediators can be probed by radiative vector meson decays. In the minimal case of a Higgs-mixed scalar, the mediator couples to quarks proportionally to their mass, and thus our signatures are not competitive with collider and $B$ meson constraints~\cite{Krnjaic:2015mbs}, which already exclude most of the thermal relic target. On the other hand, it may be possible to probe new parameter space in models where the scalar preferentially couples to light quarks~\cite{Batell:2017kty,Batell:2018fqo,Batell:2021xsi}.

Axial vector mediators can lead to invisible pseudoscalar and radiative vector meson decay, and in minimal models, they couple universally to up-type and down-type quarks~\cite{Kahn:2016vjr}. However, axial vectors are very strongly constrained through the FCNC processes $K \to \pi A'$ and $B \to K A'$, which occur with $1/m_{A'}^2$ enhanced rates due to the vector's coupling to a nonconserved current~\cite{Dror:2017nsg}. This will be the strongest constraint at low $m_{A'}$, but our meson decays may be competitive for $A'$ masses above the $K \to \pi A'$ threshold. A similar story applies for pseudoscalar mediators~\cite{Dolan:2014ska}, as they behave like the longitudinal components of light axial vectors. 

Our results could also be generalized by changing the type of DM considered. We have focused on pseudo-Dirac DM because it is a simple option consistent with all cosmological constraints. It is also a conservative choice, since its thermal freeze-out region is the most difficult to probe. However, we do not expect any of our results to qualitatively depend on this choice. The coupling of the mediator to quarks crucially determines which meson decays are allowed because the initial meson states have definite $P$ and $C$. But since the DM is generally produced at least semi-relativistically, it does not necessarily exit in the $s$-wave, which implies that the particular coupling of the mediator to DM is not as important. Therefore, we expect similar results to apply for scalar and Majorana DM, except that the falloff of the sensitivity near threshold, $m_V \approx 2 m_\chi$, may differ. For this reason, we have not shown constraints from DM direct detection, which depend sensitively on the type of DM. 

Finally, new physics that explicitly violates flavor could give rise to invisible or radiative neutral kaon decays~\cite{Gninenko:2015mea,Barducci:2018rlx}, but this must compete with stringent existing flavor constraints, such as from $K^+$ decays. For instance, in a sterile neutrino model, these constraints imply $\Br(K_L \to \nu \nu) \lesssim 10^{-10}$~\cite{Abada:2016plb}, which could possibly be probed by dedicated searches, but is likely out of reach of the strategy described in this paper. On a related note, one can consider models where the dark sector particles carry baryon number, which can then lead to baryon decays with missing energy~\cite{Heeck:2020nbq}, such as the invisible decay of neutral hyperons~\cite{Alonso-Alvarez:2021oaj}. At NA64, this can appear as a missing energy signal if the hyperon is produced by, e.g.~$\gamma p \to K^+ \Lambda$ at large momentum transfer, so that it carries most of the photon's energy. However, the dark sector particle masses must fall within a narrow window so that the analogous nucleon decays are kinematically forbidden, to avoid much stronger constraints on proton and neutron decays, and there are again potentially strong but model-dependent flavor constraints. 

\subsection{Neutrino Constraints}
\label{subsec:neutrino}

Independent of dark matter, the signatures discussed here can be used to test any model that enhances meson decays with neutrinos in the final state. For example, a new gauge boson that couples to both quarks and neutrinos can mediate invisible vector meson decay or radiative pseudoscalar meson decay. Assuming the quarks and neutrinos have comparable charges, the latter process is likely more sensitive since it is suppressed by only $g^2$, where $g$ is the gauge coupling, while the former is suppressed by $g^4$. 

Many of the strongest constraints on new light gauge bosons, such as electron beam dump experiments, rely on the coupling to electrons. Thus, light gauge bosons that couple to a combination of $B$, $L_\mu$, and $L_\tau$~\cite{Tulin:2014tya,Dobrescu:2014fca,Ilten:2018crw,Bauer:2018onh} are subject to fewer constraints. As a concrete example, a light $B - 3 L_\tau$ gauge boson would be nonanomalous, assuming the introduction of a right-handed neutrino, and predominantly decay invisibly to $\tau$ neutrinos. The leading direct constraints on such a particle are largely from measurements of radiative pseudoscalar decays~\cite{Kling:2020iar}, which, as mentioned in section~\ref{subsec:add_meson}, could be significantly improved upon by missing energy experiments. 

However, there are also much stronger indirect constraints on such gauge bosons, due to constraints on neutrino non-standard interactions from measurements of neutrino oscillations~\cite{Heeck:2018nzc,Han:2019zkz}. Therefore it may be more interesting to take a more model-independent, effective field theory point of view. As shown in Ref.~\cite{Li:2020lba}, a number of four-fermion operators involving strange quarks or tau neutrinos are unconstrained by measurements of neutrino oscillations or the CE$\nu$NS process. For these operators, the best existing constraints come from the relatively weak bounds on invisible vector meson decay, and the potential $5$ order of magnitude improvement of these bounds at LDMX would resolve this blind spot. 

\subsection{Experimental Prospects}

The potential of the invisible meson decay signal immediately suggests several avenues of further study, to refine our rough estimates. As a first step, Monte Carlo simulations could be used to compute the flux and spectrum of Bremsstrahlung photons, account for the detailed composition of the front of the calorimeters where photoproduction dominantly occurs, and to better understand the detector (non-)response at NA64 and LDMX to the recoil energy left behind. The optical models used for the meson yields, which we estimate have uncertainties ranging from 25\% to 50\%, could be substantially improved by dedicated photoproduction measurements on nuclei at the relevant energies. Ideally, however, one would additionally perform in situ measurements to assess the meson yield and experimental efficiency, such as by ``removing'' tracks from meson production and decay in real events. At LDMX, it may also be possible to measure mesons produced in the target itself. 

Meson production could also be considered as an explicit factor in the LDMX experimental design. For example, because of the $A/Z^2$ scaling of the photoproduction probability $p_V$, the meson yield could be significantly enhanced using a preshower primarily composed of light elements. However, this would also increase photonuclear backgrounds, presenting a tradeoff against the reach from $A'$ Bremsstrahlung. In addition, since mesons can also be photoproduced in the target itself, the meson yield could be enhanced by, e.g.~replacing the tungsten target with a thicker titanium target.

Further study of vector meson production and decay is highly motivated for a number of experiments. At electron beam dumps such as E137 or the proposed ILC beam dump~\cite{Asai:2021ehn}, the exclusive photoproduction processes described here account for only a small fraction of the mesons produced. However, these mesons are highly energetic, leading to forward boosted dark matter that is more likely to hit a distant detector; it would thus be interesting to consider whether this effect could extend the beam dumps' reach. At NA64, there is already an opportunity to claim leading constraints on the invisible decays of $\rho$, $\omega$, $\phi$, and $J/\psi$, if the experimental and theoretical uncertainties can be accurately quantified. Finally, at future runs of NA64 and at LDMX, exploring this signature is essential to assessing the ultimate sensitivity to dark matter.

\begin{acknowledgements}
We thank Maxim Pospelov for encouraging studies of rare meson decays in LDMX, and Emrys Peets for collaboration in the early stages of this work. We thank Asher Berlin, Tom Eichlersmith, Sebastian Ellis, Omar Moreno, Michael Peskin, Kye Wayne Shi, Yu-Dai Tsai, and the LDMX collaboration for helpful discussions. KZ is supported by the NSF GRFP under grant DGE-1656518. 
\end{acknowledgements}

\newpage
\appendix
\section{Optical Model for Meson Photoproduction}
\label{sec:optical}

\begin{figure}[b]
\includegraphics[width=\columnwidth]{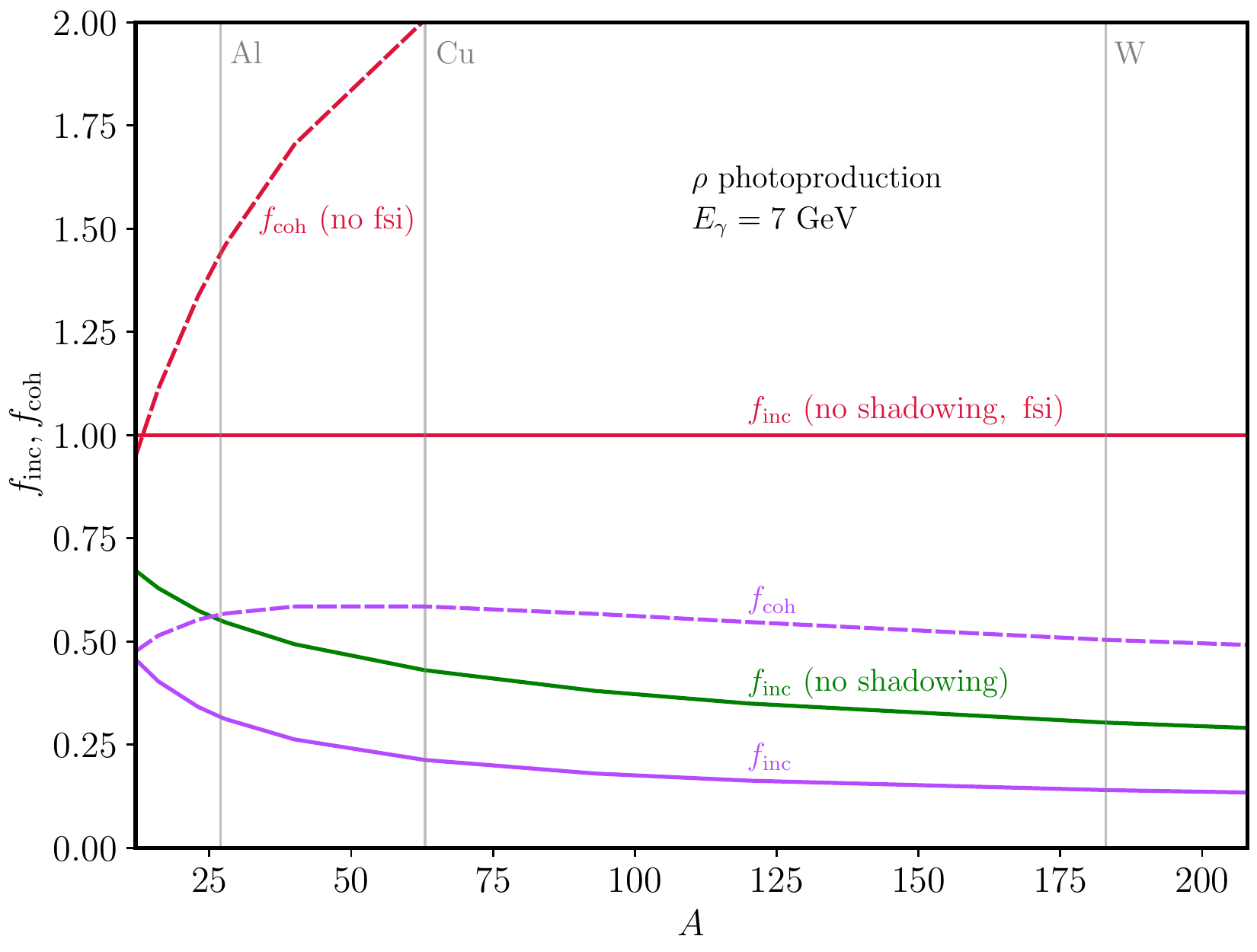}
\caption{Illustration of the effects of parts of Eqs.~\eqref{eq:coherent_prod} and~\eqref{eq:incoherent_prod}, with full results in purple. Shadowing and absorptive final state interactions play comparable roles in reducing the incoherent cross section. Without final state interactions, the coherent cross section would be much larger, with $f_{\text{coh}} \sim A^{1/3}$.}
\label{fig:R_variation}
\end{figure}

\begin{figure*}
\includegraphics[width=\columnwidth]{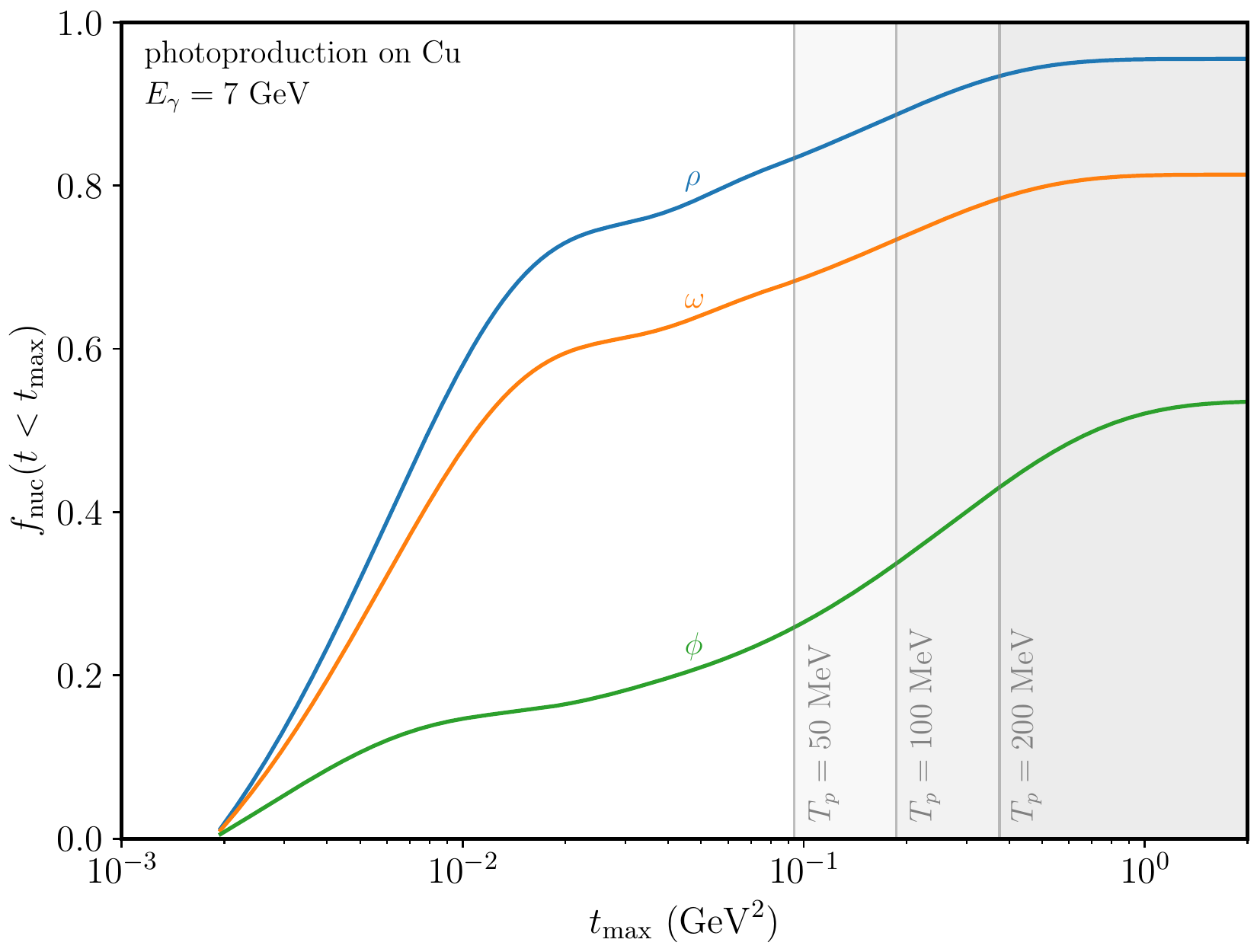}
\includegraphics[width=\columnwidth]{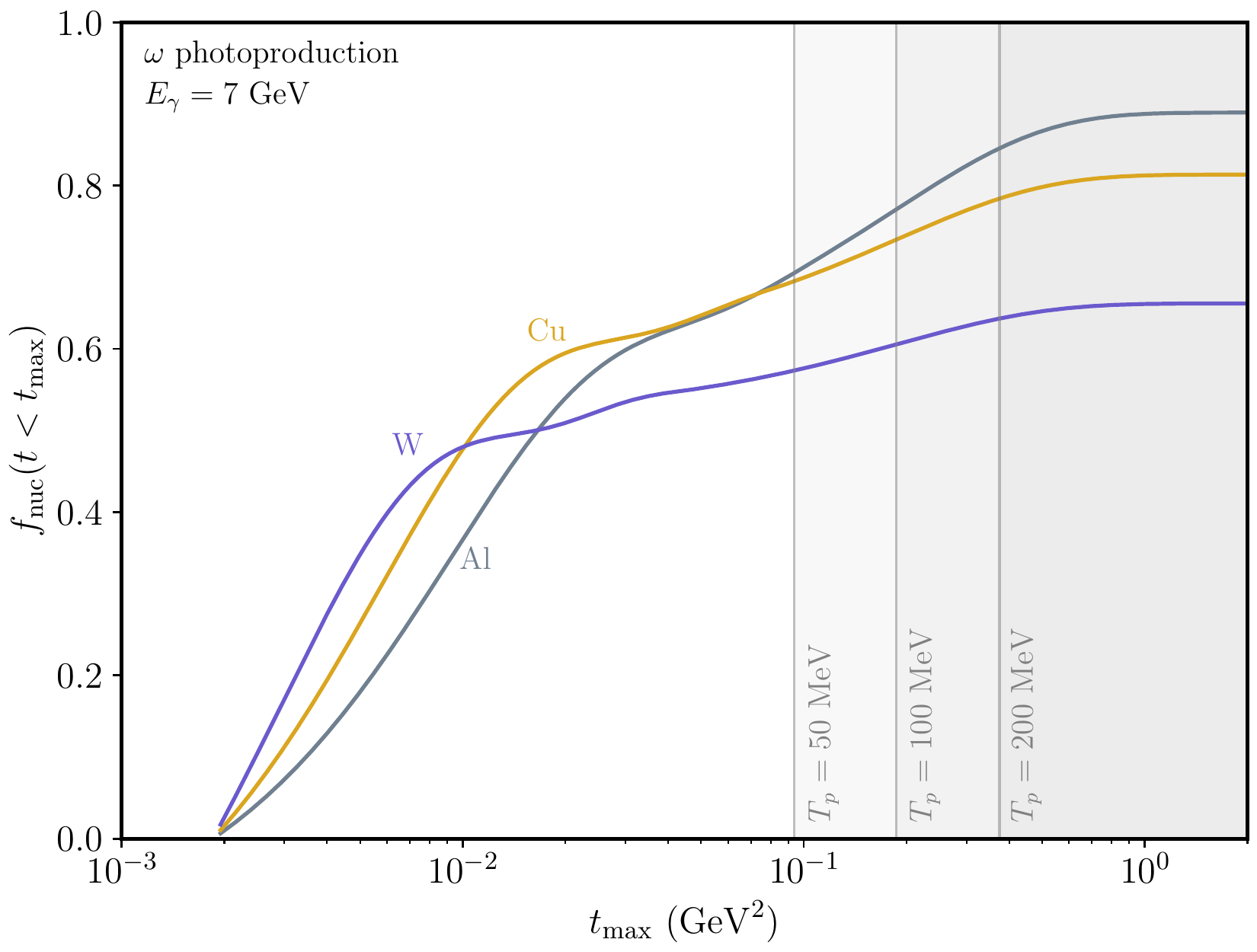}
\caption{Illustration of the integrated differential cross section for meson production, for several different mesons (left) and nuclei (right), at LDMX Phase II energies. Coherent production is sharply peaked at low $t$, especially for heavier nuclei. Incoherent production extends significantly higher in $t$, and we show vertical contours of the associated nucleon recoil energy.}
\label{fig:t_cuts}
\end{figure*}

As discussed in section~\ref{sec:photoprod}, exclusive photoproduction includes both a coherent process, where the nucleus remains in the ground state and recoils as a whole, and an incoherent process, representing the effect of photoproduction off individual nucleons. Both of these processes are well described by a Glauber optical model. Such models have a long history; we will follow the notation of the most comprehensive review~\cite{Bauer:1977iq}, but its results do not differ significantly from the first theoretical expressions written over fifty years ago, e.g.~see Ref.~\cite{Kolbig:1968rm}. First, the coherent differential cross section for photoproduction of a vector meson $V$ is
\begin{widetext}
\begin{equation} \label{eq:coherent_prod}
\frac{d\sigma_c}{dt} = \frac{d\sigma_0}{dt} \bigg|_{\theta = 0} \ \bigg| \int d^2 b \, dz \, e^{i (\mathbf{q}_T \cdot \mathbf{b} + q_\parallel z)} n(b, z) \exp\left(- \frac{\sigma_V}{2} (1 - i \alpha_V) \int_z^\infty dz' \, n(b, z') \right) \bigg|^2
\end{equation}
\end{widetext}
where $(d \sigma_0/dt)|_{\theta = 0}$ is the forward differential cross section for photoproduction of $V$ off a single nucleon, $n(b, z)$ is the nucleon number density, $\mathbf{q}_T$ and $q_\parallel$ are the transverse and longitudinal momentum transfer, $\sigma_V$ is the scattering cross section of $V$ on nucleons, and $\alpha_V$ is the ratio of the real to the imaginary part of the $V$-nucleon scattering amplitude. The first exponential factor represents the difference in phases due to photoproduction at different points in the nucleus, while the final factor accounts for absorption as the meson leaves the nucleus. 

These absorptive final state interactions have a significant effect, as shown in Fig.~\ref{fig:R_variation}. Of course, in reality, final state interactions actually produce other particles, and Monte Carlo simulations indicate that a substantial fraction of the inclusive meson yield arises from ``sidefeeding'', i.e.~production in these secondary reactions~\cite{Buss:2011mx}. However, for missing energy experiments we require a hard vector meson to carry the vast majority of the photon energy, so we are primarily interested in exclusive meson production; it is thus reasonable to treat the final state interactions as purely absorptive. 

The coherent cross section is dominated at very low momentum transfer, $|t| \ll m_p^2$, and in this regime $t \approx - (q_\parallel^2 + q_T^2)$, where $q_\parallel \approx m_V^2 / 2 E_\gamma$ is the minimum momentum transfer. In the high energy limit $q_\parallel r_{\text{nuc}} \ll 1$, the coherent peak extends to $q_T \sim 1/r_{\text{nuc}}$, leading to the narrow coherent peaks shown in Fig.~\ref{fig:t_cuts}. For heavy nuclei at few-$\GeV$ energies, we have $q_\parallel r_{\text{nuc}} \gtrsim 1$, which leads to the suppressed coherent cross sections shown in Fig.~\ref{fig:R_mass}. 

It is interesting to compare these results to the cross section for $A'$ Bremsstrahlung, which can be estimated in the Weizs{\"a}cker--Williams approximation (e.g.~see Ref.~\cite{Bjorken:2009mm}). The integral in Eq.~\eqref{eq:coherent_prod} effectively defines a form factor qualitatively similar to the elastic nuclear form factor in $A'$ Bremsstrahlung, with both falling off for $t \gtrsim 1/r_{\text{nuc}}^2$. However, $A'$ Bremsstrahlung is a $2 \to 3$ process with an intermediate virtual photon, leading to a differential cross section additionally weighted by $(t - t_{\text{min}})/t^2$ which softens the dependence on $r_{\text{nuc}}$. This is responsible for the $Z^2$ scaling of $A'$ Bremsstrahlung at low $m_{A'}$, in contrast to the rough $A^{4/3}$ scaling for coherent photoproduction. 

Because $t_{\text{min}} \sim (m_{A'}^2 / 2 E_e)^2$ for $A'$ Bremsstrahlung, the falloff in $A'$ Bremsstrahlung for high $m_{A'}$ is similar to the falloff in coherent photoproduction for high $m_V$. However, the absorptive term in Eq.~\eqref{eq:coherent_prod} implies that not all nucleons in a heavy nucleus effectively contribute to coherent photoproduction. This effectively lowers the nuclear radius by an order-one factor, which is the reason the meson decay signature reaches somewhat higher in $m_{A'}$ than $A'$ Bremsstrahlung before being significantly form factor suppressed. For example, at LDMX Phase II, the finite beam energy suppresses coherent $\phi$ photoproduction by a factor of $\sim 4$, but it suppresses $A'$ Bremsstrahlung at $m_{A'} = m_\phi$ by over two orders of magnitude. 

Next, the incoherent differential cross section is
\begin{widetext}
\begin{multline} \label{eq:incoherent_prod}
\frac{d\sigma_i}{dt} = \frac{d\sigma_0}{dt} \, \int d^2b \, dz \, n(b, z) \exp\left( - \sigma_V \int_z^\infty dz' \, n(b, z') \right) \\ \times \bigg| 1 - \int_{-\infty}^z dz'' \, n(b, z'') \frac{\sigma_V}{2} (1 - i \alpha_V) e^{i q_\parallel (z''-z)} \exp\left(- \frac{\sigma_V}{2} (1 - i \alpha_V) \int_{z''}^z dz''' \, n(b, z''') \right) \bigg|^2.
\end{multline}
\end{widetext}
Here, the first factor accounts for absorption, while the final factor is a ``shadowing'' correction accounting for destructive interference between photoproduction at $z$, and photoproduction at $z''$ followed by scattering at $z$. (This equation has a misprint in Ref.~\cite{Bauer:1977iq}, which we have corrected.) Both of these effects are comparably important for heavy nuclei, as shown in Fig.~\ref{fig:R_variation}. Because shadowing is a coherent effect, it becomes more important at high energies, leading to the decrease of $f_{\text{inc}}$ with increasing energy shown in Fig.~\ref{fig:R_mass}. Measurements of the angular distribution $d \sigma_0/dt$ are well-described by exponentials $e^{-B|t|}$, and following the most recent measurements, we take $B = 6.4 \ \GeV^{-2}$ for $\rho$~\cite{CLAS:2001zxv}, $B = 5.4 \ \GeV^{-2}$ for $\omega$~\cite{CLAS:2002cdi}, and $B = 3.0 \ \GeV^{-2}$ for $\phi$~\cite{CLAS:2013jlg,Dey:2014tfa}. For the $J/\psi$, we take $B = 4.7 \ \GeV^{-2}$ as measured by HERA~\cite{H1:2000kis}. These quantities determine the differential cross section at high momentum transfer, shown in Fig.~\ref{fig:t_cuts}.

As discussed in section~\ref{sec:expt}, we demand a nucleon recoil energy $T_p \leq 100 \ \MeV$ for incoherent photoproduction, but the precise veto efficiency is somewhat uncertain. However, because the LDMX and NA64 calorimeters are dominantly comprised of heavy nuclei, where coherent photoproduction generally dominates, the choice of cutoff does not qualitatively affect our results. Varying the cutoff by a factor of $2$ affects the light meson yields at NA64 by less than 5\%, while at LDMX it affects the $\rho$ and $\omega$ yields at the 10\% level. Coherent photoproduction is suppressed for $\phi$ at LDMX and $J/\psi$ at NA64, where the choice of cutoff leads to a uncertainty of up to 50\%.

For the parameters $\sigma_V$ and $\alpha_V$, we adopt the values of Model I of Ref.~\cite{Bauer:1977iq}, which were motivated by quark model estimates and chosen to adequately describe coherent photoproduction data. The results do not depend strongly on the nuclear density model, but we use a Woods--Saxon distribution~\cite{Chudakov:2015msa}, 
\begin{equation}
n(r) \propto \frac{1}{1 + e^{(r-c)/a}}
\end{equation}
where $c = 1.12 \, A^{1/3} \, \text{fm}$ and $a = 0.545 \, \text{fm}$. We fix the total single nucleon photoproduction cross sections $\sigma_0$ to data, as described in section~\ref{sec:photoprod}. 

Because coherent production is so sharply forward peaked, the coherent and incoherent channels may be measured separately by placing restrictions on $t$. As reviewed in Ref.~\cite{Bauer:1977iq}, coherent $\rho$ and $\omega$ photoproduction have been thoroughly measured for a variety of nuclei and photon energies, and are well-described by optical models to within an uncertainty of at most 25\%. Coherent $\phi$ photoproduction is less well-measured; the data can still be fit, but with a larger uncertainty in $\alpha_\phi$ and $\sigma_\phi$. 

There is much less data available for incoherent photoproduction, particularly at high photon energies, and the data that exists is more ambiguous. At the time of writing of Ref.~\cite{Bauer:1977iq}, the data for incoherent $\rho$ photoproduction was sufficient to confirm the existence of a shadowing effect, but not enough to investigate it in detail. We have chosen to implement the simplest version of it, but theoretically reasonable modifications of Eq.~\eqref{eq:incoherent_prod} could change the cross section by as much as 50\%, while still fitting the data comparably well. 

More recently, a number of experiments have measured incoherent photoproduction on nuclei, motivated by anomalous results for $\phi$ mesons at SPring-8~\cite{Ishikawa:2004id} (for a recent review, see Ref.~\cite{CLAS:2010pxs}). The measurements indicate a steep falloff of $f_{\text{inc}}$ with increasing $A$ for low photon energies, corresponding to an absorption cross section $\sigma_\phi$ dramatically above the quark model expectation. Several theoretical works have proposed explanations based on ``in-medium'' modifications of the $\phi$ meson width, while Ref.~\cite{Sibirtsev:2006yk} considers the alternative of $\omega$ to $\phi$ transitions. Currently, these puzzling results do not seem to have a canonical explanation, and different experiments are not fully in agreement; thus, we regard our estimate of $\phi$ production at LDMX Phase I to be uncertain within a factor of $2$. Fortunately, these in-medium effects should become less important at the higher energies of LDMX Phase II and NA64, where incoherent photoproduction is in any case subdominant. 

\bibliographystyle{utphys}
\bibliography{Mesons}

\end{document}